\RequirePackage{fix-cm}
\documentclass[times,doublespace]{nlaauth0} 
\usepackage{moreverb}
\usepackage[dvips,colorlinks,bookmarksopen,bookmarksnumbered,citecolor=red,urlcolor=red]{hyperref}
\usepackage{comment}
\newcommand\BibTeX{{\rmfamily B\kern-.05em \textsc{i\kern-.025em b}\kern-.08em
T\kern-.1667em\lower.7ex\hbox{E}\kern-.125emX}}

\def\volumeyear{2012}
\usepackage{url}
\usepackage[normalem]{ulem}
\usepackage{xspace}
\usepackage{graphicx}
\usepackage{boxedminipage}
\usepackage{algorithm}
\usepackage{algorithmic}
\usepackage{verbatim}
\begin{document}
\runningheads{A. Abdel-Rehim et. al.}{Extending eigCG to nonsymmetric systems}

\title{Extending the eigCG algorithm to nonsymmetric Lanczos for linear systems with multiple right-hand sides}

\author{A. M. Abdel-Rehim \affil{1},
Andreas Stathopoulos \affil{2,}\corrauth,
Kostas Orginos \affil{3,}\affil{4}}
\address{
\affilnum{1} Computation-based Science and Technology Research Center (CaSToRC), The Cyprus Institute, 
	20 Kostantinou Kavafi Street, 2121 Aglantzia, Nicosia, Cyprus \break
\affilnum{2} Department of Computer Science, College of William and Mary, Williamsburg, Virginia 23187-8795, U.S.A. \break
\affilnum{3}Department of Physics, College of William and Mary, Williamsburg, Virginia 23187-8795, U.S.A.\break
\affilnum{4}Jefferson National Laboratory, 12000 Jefferson Avenue, Newport News, Virginia, 23606, U.S.A.}

\corraddr{Department of Computer Science, College of William and Mary, Williamsburg, Virginia 23187-8795, U.S.A. E-mail: andreas@cs.wm.edu}

\renewcommand{\thefootnote}{\arabic{footnote}}
\newcommand{\cg}{{\em CG}\xspace}
\newcommand{\lanczos}{{\em Lanczos}\xspace}
\newcommand{\eigcg}{{\em eigCG}\xspace}
\newcommand{\increigcg}{{\em Incremental eigCG}\xspace}
\newcommand{\eigbicg}{{\em eigBiCG}\xspace}
\newcommand{\increigbicg}{{\em Incremental eigBiCG}\xspace}
\newcommand{\bicgstab}{{\em BiCGStab}\xspace}
\newcommand{\bicg}{{\em BiCG}\xspace}
\newcommand{\bilanczos}{{\em Bi-Lanczos}\xspace}
\newcommand{\eigbicgkm}[2]{{\em eigBiCG(#1,#2)}\xspace}
\newcommand{\gmres}{{\em GMRES}\xspace}
\newcommand{\ortheigbilanczos}{{\em biortho-eigBiCG}\xspace}
\newcommand{\Vm}{ V^{(m)} }
\newcommand{\Vmdagger}{ V^{(m)\dagger} }
\newcommand{\Wm}{ W^{(m)} }
\newcommand{\Wmdagger}{ W^{(m)\dagger} }
\newcommand{\Tm}{ T^{(m)} }
\newcommand{\normx}[1]{||#1||}
\newcommand{\cblue}[1]{{\color{black}#1}}
\newcommand{\cred}[1]{{\color{red}#1}}
\newcommand{\initbicg}{{\em init-BiCG}\xspace}
\newcommand{\initbicgstab}{{\em init-BiCGStab}\xspace}
\begin{abstract}
The technique that was used to build the \eigcg algorithm for sparse symmetric linear systems is extended to the 
nonsymmetric case using the \bicg algorithm. We show that, similarly to the symmetric case, we can 
build an algorithm that is capable of computing a few smallest magnitude eigenvalues and their 
corresponding left and right eigenvectors of a nonsymmetric matrix using only a small window of the 
\bicg residuals while simultaneously solving a linear system with that matrix. 
For a system with multiple right-hand sides, we give an algorithm that computes incrementally more eigenvalues
while solving the first few systems and then uses the computed eigenvectors to deflate \bicgstab for the 
remaining systems. 
Our experiments on various test problems, including Lattice QCD, show the remarkable ability of \eigbicg 
to compute spectral approximations with accuracy comparable to that of the unrestarted, nonsymmetric Lanczos.
Furthermore, our incremental \eigbicg followed by appropriately restarted and deflated \bicgstab provides 
a competitive method for systems with multiple right-hand sides. 
\end{abstract}
\pagestyle{myheadings}
\thispagestyle{plain}
\maketitle
\vspace{-6pt}
\section{Introduction}
Many scientific and engineering applications require the solution of linear systems of equations with many right-hand sides $b_i$:
\begin{equation}
Ax_i=b_i,  \quad \quad i=1,2,\dots,n_s,
\label{main-problem}
\end{equation}
where $A$ is a large, sparse, nonsymmetric matrix of dimension $n$. 
Efficient algorithms should take advantage of the fact that all these systems correspond to the same matrix. 
Because of size and sparsity, dense-matrix methods that reuse the matrix factorization cannot be used. 
Krylov iterative methods 
\cite{Saad:2003,Simoncini_Szyld_RecentDevelopments:2007} are the fundamental tool to solve such systems.
However, they build a separate iteration for each system and, thus,
can be inefficient, especially when the number of right-hand sides is large. 
Variants of Krylov methods that exploit the common matrix on multiple right 
hand sides have been proposed in the 
literature. These include block methods \cite{Saad:2003,Gutknecht_block_intro, Golub:1977, 
OLeary:1980, ElGuennouni:2004, blockQMR, blockGMRES, blockseedBiCGStab, blockBiCGStab},
seed methods \cite{Smith:Seed:PhD:1987, Smith:Seed:1989, Simoncini:1995, Chan:1997, Saad:Seed:1987}, 
deflation methods \cite{initCG, Morgan:2002, Wang_Sturler_Paulino, Ahuja:Recycled_bicg:2010, TRLAN, Stathopoulos:1998, eigPCG, lan-dr, Morgan:2008},  
and their combinations \cite{restarted_block_gmres_Morgan:2005,Misha_seed}. 
We focus on deflation methods as they do not require all the right-hand sides to be available from the start 
(as block methods do) and extract intrinsic information about the common matrix, not in relation to the right hand sides 
(as seed methods do).

Deflation is based on the fact that, for a large class of ill conditioned 
problems, the slow convergence of Krylov linear system solvers
is caused by small eigenvalues of the matrix $A$. If the eigenvectors corresponding to those small eigenvalues were known, 
one could project them out (deflate them) from the initial residual and then solve the deflated system, which will converge 
much faster. 
Although other issues relating to eigenvalue distribution and 
conditioning may also cause problems to nonsymmetric Krylov methods, 
for many applications the problem is in the small eigenvalues, and where 
most current deflation research focuses.
Moreover, preconditioners are often used to deal with these other issues, and 
deflation can applied on the preconditioned matrix for further improvements.

In principle, one can use a 
separate eigensolver \cite{ARPACK,PRIMME} to compute small eigenvalues of $A$ and then use them to deflate (\ref{main-problem}). 
However, it is more efficient to compute the small eigenvalues simultaneously while solving 
the linear systems. Recently, we proposed an algorithm that uses such strategy for Symmetric Positive Definite (SPD) matrices 
\cite{eigPCG}. The algorithm---called \eigcg---has the following features:
\begin{enumerate}
\item The linear system is solved with the Conjugate-Gradient (CG) 
      algorithm which is computationally and memory efficient.
\item While solving the linear system, \eigcg computes a few small eigenvalues and eigenvectors 
   using only a small window of the CG residuals.
\item The computation of the eigenvalues does not affect the solution of the linear system, and no restarting
      of the linear system occurs.
\item \eigcg computes small eigenvalues with the same efficiency and almost the same accuracy as unrestarted
Lanczos, using much smaller memory requirements. 
\end{enumerate}
The number and precision of the few eigenvalues computed by \eigcg while solving a single right-hand side
are usually not sufficient for efficient deflation of subsequent systems. To compute more eigenvalues and
improve their accuracy, we developed the \increigcg algorithm. Our tests on various problems 
showed that \increigcg was able to compute accurately a large number of eigenvalues and solve systems
with multiple right hand sides with speed-ups up to an order of magnitude over undeflated CG.

The reason for the success of \eigcg can be traced to a combination of thick and locally optimal restarting
techniques for eigenvalue problems \cite{Nearly_Optimal_I,Nearly_Optimal_II,LOBPCG}.
These techniques manage to maintain appropriate orthogonality information during restarts of a search space
so that the optimality of the Galerkin procedure continues to hold as if on the unrestarted Krylov space.
What is surprising with \eigcg is that these techniques continue to work when future iteration vectors are not 
generated based on this space (as in subspace iteration) but borrowed from a Lanczos or CG process \cite{eigPCG}.

In this paper we study the extension of \eigcg to the nonsymmetric case.
Our goal is similar: approximate eigenvectors from a small search space 
that is obtained as a by-product of some Krylov method (of Arnoldi or 
\bicg type) and maintains approximately the orthogonality over all seen 
Krylov vectors. 
The subspace built by Arnoldi type methods is typically restarted,
  and thus loses global orthogonality against past vectors which
  cannot be recovered effectively with our \eigcg technique.
Other efforts to correct this have resulted in somewhat limited success
  \cite{Sturler_trunc,Alison_Baker_05}.
Therefore, we turn to the \bicg method because 
    (1) it uses an inexpensive three term recurrence to produce a biorthogonal 
        Krylov basis, at least in exact arithmetic, and 
    (2) the restarting technique used in \eigcg is effective in the context 
 	of biorthogonal eigenvalue solvers \cite{AS_biJD_2002}.

The new algorithm is called \eigbicg and computes a few eigenvalues and their corresponding left
and right eigenvectors using a small window of \bicg residuals while solving a linear system.
The \bicg method is unaffected.
For multiple right-hand sides, we extend the \increigcg to the \increigbicg algorithm. 
We first solve a few systems accumulating eigenvectors with \increigbicg. 
Using these eigenvectors, the rest of the systems are solved by 
deflated \bicgstab, which can especially benefit from deflation with both left and 
right eigenvectors \cite{Frank:2001}.

For the eigenvalue computation phase, we use \bicg instead 
of \bicgstab because the Lanczos parameters and space are readily available
in \bicg.
Recently, it has been shown that Ritz values and right Ritz vectors could be 
computed using the $IDR$ algorithm, which is related to \bicgstab 
\cite{Zemke:Doshisha:2011}. 
Such a method might solve the initial few linear systems a little more 
efficiently than \bicg, but it would incur additional costs to find the 
eigenvectors.
More importantly, it is not clear how to obtain the left eigenvector space
from \bicgstab.
Either way, the majority of the systems are already solved with 
  deflated \bicgstab, so exploring this potential method is beyond 
  the scope of the current paper.

There are other algorithms in the literature for solving systems with multiple right-hand sides using deflation.
We mention in particular Lanczos with deflated restarting (Lan-DR) \cite{lan-dr,Morgan_LANDR2},
{\it GMRes} with deflated restarting ({\it GMRes-DR} and {\it GMRes-Proj}) for the nonsymmetric case 
\cite{Morgan:2002,Morgan:2004,Morgan:2008}, and Recycled Krylov methods \cite{Wang_Sturler_Paulino,Ahuja:Recycled_bicg:2010}. 
The algorithms we propose are different in several ways.
{\it GMRes} type algorithms solve both the linear system and eigenvalue problem with restarted Arnoldi 
while \eigbicg solves the linear system with an unrestarted method. 
Although our eigenvector search space is restarted, 
our experiments show that convergence is similar to the unrestarted bi-Lanczos. 
In some cases, this yields better eigenvalue approximations than the restarted Arnoldi. 
Also, {\it GMRes-DR} obtains the eigenvectors from a single linear system and does not update them subsequently.
Recycled \bicg is closer to \eigbicg as it is a two sided method and uses a small
eigenvector search space borrowed from unrestarted \bicg. 
However, without the locally optimal restarting technique, its spectral approximations are not accurate eigenvectors
and therefore have been used mainly in applications where the matrix changes between right hand sides.
On the other hand, the deflated nonsymmetric Lanczos in \cite{Morgan_LANDR2} is a thick restarted eigensolver.
For deflation, other methods project the obtained eigenvectors at every step ({\it GMRes}, Recycled \bicg)
or at every restart ({\it GMRes-Proj}). This adds an expensive overhead when the number of eigenvectors is large.
Our methods deflate a linear system only a small, constant number of times which is independent of
the convergence of the system.

\cblue{
We want to point out at the outset an inherent limitation of all deflation methods.
For many applications, such as PDEs or our motivating application from lattice quantum chromodynamics (QCD), 
  the density of the eigenvalues near zero grows linearly with the matrix size, $n$.
Thus, to achieve a constant number of iterations with growing $n$, the cost of deflation becomes $O(n^2)$, 
  and the cost of obtaining these eigenvectors becomes $O(n^3)$.
Although the constants in the complexity are small, for a sufficient large $n$ multigrid methods should
  scale better than deflation \cite{Brandt_77}.
Recent advances in lattice QCD, in particular, have resulted in a version of algebraic multigrid where 
  the interpolators are generated by an approximate near null eigenspace \cite{Luscher:2007se, Babich:2010qb}.
Generating this preconditioner is also expensive, but researchers have started to see benefits in some of 
  the larger lattices today.
In this paper, we focus on problems that do not fall in this asymptotic realm or on problems where the
   preconditioner has not fully removed all low magnitude eigenvalues.
}
 
In the following we denote by $\bar{A}$, $A^T$, $A^\dagger$ the complex conjugate, the transpose, and the Hermitian
conjugate of a non-defective matrix $A$ respectively.  
We denote by $<w,v>=w^\dagger v$ the dot product of two vectors $v$ and $w$,
and we use $||\cdot ||$ as the 2-norm of vectors and matrices.
The complex conjugate and the norm of a complex number $\alpha$ are denoted by 
$\bar{\alpha}$ and $|\alpha|$ respectively. 
$V^{(m)}$, or $V$ when there is no ambiguity, represents a matrix whose 
columns are the vectors $v_1,v_2,\dots,v_m$.
When the number of columns is changing we use the notation $V=[v_1,v_2,\dots]$.

\section{Background}
\label{sec:background}
\subsection{Eigenvalue computation in \eigcg}
\begin{figure}[hbtp]
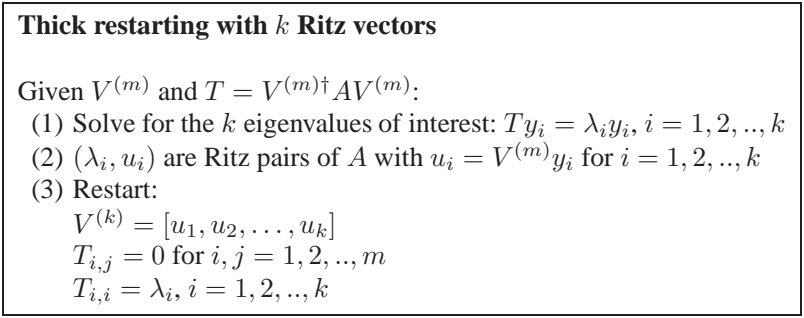

\begin{center}
\fbox{
\begin{minipage}[t]{\textwidth}
\begin{tabbing}
x\=xxx\=xxx\=xxx=\kill
{\bf Thick restarting with $k$ Ritz vectors}\\
\> \\
Given $V^{(m)}$ and $T=\Vmdagger A \Vm$:\\
\> (1) \> Solve for the $k$ eigenvalues of interest: $Ty_i = \lambda_i y_i$, $i=1,2,..,k$\\
\> (2) \> $(\lambda_i, u_i)$ are Ritz pairs of $A$ with $u_i=V^{(m)}y_i$ for $i=1,2,..,k$\\
\> (3) \> Restart: \\
\>     \> $V^{(k)}=[u_1,u_2,\dots,u_k]$\\
\>     \> $T_{i,j}=0$ for $i,j=1,2,..,m$\\
\>     \> $T_{i,i}=\lambda_i$, $i=1,2,..,k$
\end{tabbing}
\end{minipage}
}
\caption{Thick restarting with $k$ Ritz vectors: symmetric case.}
\label{fig:eig_restart_k}
\end{center}
\end{figure}

\begin{figure}[hbtp]
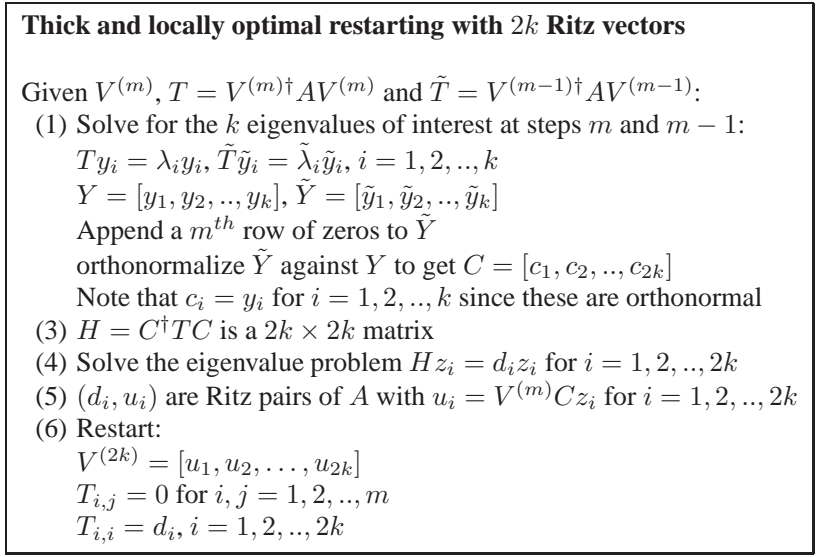

\begin{center}
\fbox{
\begin{minipage}[t]{\textwidth}
\begin{tabbing}
x\=xxx\=xxx\=xxx=\kill
{\bf Thick and locally optimal restarting with $2k$ Ritz vectors}\\
\> \\
Given $V^{(m)}$, $T=\Vmdagger A \Vm$ and $\tilde{T}=V^{(m-1)\dagger} A V^{(m-1)}$:\\
\> (1) \> Solve for the $k$ eigenvalues of interest at steps $m$ and $m-1$:\\
\>     \> $Ty_i=\lambda_i y_i$, $\tilde{T} \tilde{y}_i=\tilde{\lambda}_i \tilde{y}_i$, $i=1,2,..,k$\\
\>     \> $Y=[y_1,y_2,..,y_k]$, $\tilde{Y}=[\tilde{y}_1,\tilde{y}_2,..,\tilde{y}_k]$\\
\>     \> Append a $m^{th}$ row of zeros to $\tilde{Y}$\\
\>     \> orthonormalize $\tilde{Y}$ against $Y$ to get $C=[c_1,c_2,..,c_{2k}]$\\
\>     \> Note that $c_i=y_i$ for $i=1,2,..,k$ since these are orthonormal\\
\> (3) \> $H=C^\dagger T C$ is a $2k\times 2k $ matrix\\
\> (4) \> Solve the eigenvalue problem $Hz_i=d_i z_i$ for $i=1,2,..,2k$\\
\> (5) \> $(d_i,u_i)$ are Ritz pairs of $A$ with $u_i=V^{(m)} C z_i$ for $i=1,2,..,2k$\\
\> (6) \> Restart: \\
\>     \>  $V^{(2k)}=[u_1,u_2,\dots,u_{2k}]$\\
\>     \>  $T_{i,j}=0$ for $i,j=1,2,..,m$\\
\>     \>  $T_{i,i}= d_i$, $i=1,2,..,2k$
\end{tabbing}
\end{minipage}
}
\caption{Thick and locally optimal restarting with $2k$ Ritz vectors: symmetric case.}
\label{fig:eig_restart_2k}
\end{center}
\end{figure}

We first review how the \eigcg algorithm computes approximations 
to a few eigenvalues inside \cg using a subspace of limited size and how this subspace is restarted. 
Assume we look for $k$ smallest eigenpairs of an SPD matrix $A$ of dimension $n$.
Let $m>k$ be the maximum dimension of the subspace that will be used to
compute the approximate eigenvectors.
Denote by $V^{(m)} \in \Re^{n\times m}$ an orthonormal basis of this subspace. 
After $m$ steps of \lanczos (or \cg), $V^{(m)}$ 
holds the first $m$ Lanczos vectors (or \cg residuals properly normalized). In a plain thick restarting
approach \cite{Stathopoulos:1998,TRLAN}, we would compute $k$ Ritz vectors of interest and restart the subspace 
with these $k$ Ritz vectors (see Figure \ref{fig:eig_restart_k}). 
Then, we would continue the iteration, filling the remaining $m-k$ positions in the basis with new Lanczos vectors.
This approach is followed in Recycled {\it MINRES}
but does not approximate the eigenpairs very well \cite{Wang_Sturler_Paulino}.
In \eigcg, we restart not only with the $k$ Ritz vectors computed at step $m$, 
but also with the $k$ Ritz vectors computed at step $m-1$ (if $m > 2k$).
For stability, the $2k$ vectors are orthonormalized. 
The remaining $m-2k$ positions of the basis are then filled with new Lanczos vectors.
This approach for restarting the eigenvalue search subspace 
is based on Locally Optimal CG (LOCG) and in eigensolvers consistently yields
 convergence which is almost indistinguishable from unrestarted Lanczos 
\cite{eigPCG,DYAKONOV_83,Knyazev_91,LOBPCG,CM_SR_ED_92a,AS_YS_98,Nearly_Optimal_I,Nearly_Optimal_II}.
Surprisingly, it performs equally well when the search space 
is made of recycled Lanczos vectors.
Orthogonalization of the eigenvectors from steps $m$ and $m-1$ can be done with small vectors of length $m$ 
at negligible cost. Figure \ref{fig:eig_restart_2k} shows how this is implemented. 


\subsection{\bilanczos algorithm}
Given vectors $v_1, w_1$ with $<w_1,v_1>=1$,
$m$ iterations of the {\it Bi-Lanczos} algorithm \cite{Lanczos,Saad:2003} build biorthogonal bases 
$\Vm =[v_1,\dots,v_m]$ and 
$\Wm =[w_1,\dots,w_m]$ of the Krylov subspaces 
\begin{equation}
\begin{split}
{\cal K}_r^{(m)}(A,v_1) &= \mbox{span}\{v_1, Av_1, A^2v_1,\dots,A^{m-1}v_1\} \\
{\cal K}_l^{(m)}(A^\dagger,w_1) &= \mbox{span}\{w_1, A^\dagger w_1, {A^\dagger}^2w_1,\dots,{A^\dagger}^{m-1}w_1\}
\end{split}
\end{equation}
using a three-term recurrence with a tridiagonal projection matrix $T = \Wmdagger A \Vm$.
To solve a linear system $Ax=b$ with initial guess $x_0$, $v_1$ is chosen as 
$v_1=r_0=b-Ax_0$, and the solution is given by: $x=x_0 + \Vm T^{-1} \Wmdagger r_0$.
Using the Rayleigh-Ritz procedure on $\Vm$ and $\Wm$, we can also compute $m$ approximate eigentriplets of $A$. 
If $y$ and $z$ are right and left eigenvectors of $T$ corresponding to the eigenvalue $\lambda$, then
$p=\Vm y$ and $q=\Wm z$ are the right and left Ritz vectors of $A$ corresponding to the Ritz value $\lambda$.  
Note that in order to compute approximate eigenvectors, we need to store all the basis vectors $\Vm$ and $\Wm$ or
re-compute them. For solving a linear system, this storage is not needed as $x$ is given by the \bicg three-term recurrence.

\subsection{\bicg algorithm}
The \bicg algorithm \cite{bicg} is derived form the \bilanczos algorithm by replacing the three-term recurrence
by a coupled two-term recurrences. For solving the linear system $Ax=b$ with initial guess $x_0$, the algorithm 
is given in Figure \ref{fig:bicg-algorithm}.
The biorthogonal basis vectors $V=[v_1,v_2,\dots]$ and $W=[w_1,w_2,\dots]$ of the \bilanczos 
algorithm are parallel to the \bicg residuals as
\begin{equation}
v_{j+1}=\theta_j r_j, \quad w_{j+1} = \delta_j \hat{r}_j, \quad j=0,1,\dots.
\label{eq:wv_norms}
\end{equation}
The normalization factors $\theta_j$ and $\delta_j$ are chosen such that $<w_{j+1},v_{j+1}>=1$. 
We choose the following normalization which balances the norm of $v_{j+1}$ and $w_{j+1}$,
\begin{equation}
\theta_j = \frac{1}{\sqrt{|<\hat{r}_j,r_j>|}}=\frac{1}{\sqrt{|\rho_j|}}, \quad
\delta_j = \frac{\sqrt{|<\hat{r}_j,r_j>|}}{<r_j,\hat{r}_j>} = \frac{\sqrt{|\rho_j|}}{\bar{\rho}_j}. 
\end{equation}
The elements of the tridiagonal projection matrix $T = \Wmdagger A \Vm$ can also be computed from the scalars in 
the \bicg algorithm (see also \cite{Ahuja:Recycled_bicg:2010}). Using Equation (\ref{eq:wv_norms}), the relations
\begin{equation}
r_j = p_j-\beta_{j-1}p_{j-1}, \quad
\hat{r}_j = \hat{p}_j - \bar{\beta}_{j-1} \hat{p}_{j-1}, 
\end{equation}
and the biorthogonality conditions of the \bicg algorithm
$<\hat{p}_k,Ap_l> = 0, \ k\neq l$,
we find 
\begin{equation}
\label{eq:tridiagonal}
\begin{split}
T_{1,1}     &= \frac{1}{\alpha_0},\\ 
T_{j+1,j+1} &= \frac{1}{\alpha_{j}}+\frac{\beta_{j-1}}{\alpha_{j-1}},  \quad j=1,2,\dots, \\ 
T_{j+1,j+2} &= -\bar{\delta}_j \theta_{j+1} \beta_j <\hat{p}_j,Ap_j>,  \quad j=0,1,2,\dots, \\
T_{j+2,j+1} &= -\bar{\delta}_{j+1} \theta_j \beta_j <\hat{p}_j,Ap_j>,  \quad j=0,1,2,\dots. \\
\end{split}
\end{equation}
These relations will be useful for computing approximate eigenpairs inside \bicg.
 
\begin{figure}[htbp]
\begin{center}
\fbox{
\begin{minipage}[t]{\textwidth}
\begin{tabbing}
x\=xxx\=xxx\=xxx=\kill
{\bf The BiCG Algorithm:}\\
\> \\
Solve $Ax=b$ given initial guess $x_0$\\
\> (0) \> $r_0=b-Ax_0$, $p_0=r_0$\\
\>     \> Choose $\hat{r}_0$ such that $<\hat{r}_0,r_0> \neq 0$\\
\>     \> $\hat{p}_0=\hat{r}_0$, $\beta_{-1}=0$\\
\>     \> $\rho_0=<\hat{r}_0,r_0>$, if $\rho_0=0$ stop\\
\> (1) \> for  $j=0,1,2,\dots$ till convergence\\
\> (2)  \>\> $\alpha_j = \rho_j / <\hat{p}_j,Ap_j>$\\
\> (3)  \>\> $x_{j+1} = x_j + \alpha_jp_j$\\
\> (4)  \>\> $r_{j+1} = r_j -\alpha_j A p_j$\\
\> (5)  \>\> $\hat{r}_{j+1} = \hat{r}_j - \bar{\alpha}_j A^\dagger \hat{p}_j$\\
\> (6)  \>\> $\rho_{j+1}=<\hat{r}_{j+1},r_{j+1}>$, if $\rho_{j+1}=0$ stop\\ 
\> (7)  \>\> $\beta_j=\rho_{j+1}/\rho_j$\\
\> (8)  \>\> $p_{j+1} = r_{j+1} + \beta_j p_j$\\
\> (9)  \>\> $\hat{p}_{j+1} = \hat{r}_{j+1} + \bar{\beta}_j \hat{p}_j$ 
\end{tabbing}
\end{minipage}
}
\caption{The \bicg algorithm for solving a linear system $Ax=b$}
\label{fig:bicg-algorithm}
\end{center}
\end{figure}

\section{The eigBiCG Algorithm}
\label{sec:eigbicg}
We augment the standard \bicg algorithm with a 
part that approximates a few eigentriplets using the \bicg residuals, 
$\Vm, \Wm$, which we restart similarly to \eigcg (Figure \ref{fig:eig_restart_2k}). 
The difference is that in \eigbicg we deal with two biorthogonal bases. 
In \cite{AS_biJD_2002}, we suggested such a restarting approach in the 
context of a biorthogonal Jacobi-Davidson (JD) method. As with linear systems, 
restarting causes a slowdown in convergence of eigensolvers. Moreover,
in the nonsymmetric case, certain Ritz values may cease to converge or
disappear completely from the restarted basis. When the left and right 
eigenspace is not too ill-conditioned, our technique managed to 
alleviate and sometimes eliminate these effects.
The difference between \eigbicg and JD is that the restarted eigenvalue 
  search space is not used to determine subsequent iteration vectors.
For the same reason, restarting
has no effect on the solution of the linear system.  

\subsection{Computing eigenvalues and eigenvectors in \bicg}
\label{subsec:comp_evals_evecs_bicg}
Let $k$ be the number of eigenpairs we need to compute, for example those with smallest absolute value, 
and $m$ be the size of the right and left subspaces $\Vm$ and $\Wm$ such that $m>2k$. We
compute $2k$ approximate Ritz vectors and values (from steps $m-1$ and $m$) and restart $\Vm$ and $\Wm$ as
shown in Figure \ref{fig:bicg_eigpart}. 

\begin{figure}[htbp]
\begin{center}
\fbox{
\begin{minipage}[t]{\textwidth}
\begin{tabbing}
x\=xxx\=xxx\=xxx=\kill
{\bf Restarting with $2k$ Ritz vectors: \bicg case}\\
\> \\
Given $\Vm$, $\Wm$ , $T=\Wmdagger A \Vm$ and $\tilde{T}=W^{(m-1)\dagger} A V^{(m-1)}$:\\
\> (1) \> Solve for the $i=1,\ldots,k$ eigentriplets of interest at steps $m$ and $m-1$:\\
\>     \> Compute $(\lambda_i, y_i, z_i)$ eigenvalues, right and left eigenvectors of $T$\\
\>     \> Compute $(\tilde{\lambda}_i, \tilde{y}_i, \tilde{z}_i)$ eigenvalues, right and left eigenvectors of $\tilde{T}$\\
\> (2) \> $Y=[y_1,y_2,..,y_k]$, $\tilde{Y}=[\tilde{y}_1,\tilde{y}_2,..,\tilde{y}_k]$\\
\>     \> $Z=[z_1,z_2,..,z_k]$, $\tilde{Z}=[\tilde{z}_1,\tilde{z}_2,..,\tilde{z}_k]$\\
\>     \> Append a $m^{th}$ row of zeros to $\tilde{Y}$, and $\tilde{Z}$\\
\> (3) \> Biorthogonalize $(\tilde{Y},\tilde{Z})$ against $(Y,Z)$ to get $(C,D)$\\
\>     \> $C=[c_1,c_2,..,c_{2k}]$ and $D=[d_1,d_2,..,d_{2k}]$\\
\>     \> Note that $c_i=y_i$ and $d_i=z_i$, $i=1,2,..,k$ since these are biorthogonal\\
\> (4) \> $H=D^\dagger \Tm C$, a $2k\times 2k $ matrix\\
\> (5) \> Compute the $2k$ eigenvalues $\gamma_i$ and the corresponding \\
\>     \> right and left eigenvectors $f_i$ and $g_i$ of $H$\\
\> (6) \> $\gamma_i,u_i,q_i$ are Ritz values, right, and left Ritz vectors of $A$ with \\
\>     \> $u_i=V^{(m)} C f_i$ and $q_i=W^{(m)} D g_i,\ i=1,2,..,2k$\\
\> (7) \> Restart: \\
\>     \>  $V^{(2k)}=[u_1,u_2,\dots,u_{2k}]$\\
\>     \>  $W^{(2k)}=[q_1,q_2,\dots,q_{2k}]$\\
\>     \>  $T_{i,j}=0$ for $i,j=1,2,..,m$\\
\>     \>  $T_{i,i}= \gamma_i$, $i=1,2,..,2k$
\end{tabbing}
\end{minipage}
}
\caption{Restarting with $2k$ Ritz vectors: nonsymmetric case.}
\label{fig:bicg_eigpart}
\end{center}
\end{figure}

After the first $m$ steps of \bicg, the bases $\Vm$ and $\Wm$ are given by the \bicg residuals 
and the projection matrix $T=\Wmdagger A \Vm$ is tridiagonal. After restarting, $T$ has
a diagonal $2k\times 2k$ block and the first $2k$ basis vectors in $\Vm$ and $\Wm$ are the 
approximate right and left Ritz vectors. 
\cblue{ Subsequent residuals from the original \bicg, 
  $r_{m+1},\hat r_{m+1}, r_{m+2}, \hat r_{m+2}, \ldots$ will be appended to the remaining $m-2k$ positions
of $V, W$, i.e., $v_{2k+1},w_{2k+1},v_{2k+2},w_{2k+2},\ldots$. 
By construction, the new residuals remain biorthogonal to all the vectors already in $V, W$, 
  and the coefficients of the tridiagonal projection matrix are computed using the equations
  in (\ref{eq:tridiagonal}).
The only exception is the vectors $v_{2k+1}$ and  $w_{2k+1}$ which need special attention.

After restarting, 
the elements $T_{i,2k+1}=q_i^\dagger A v_{2k+1}$ and $T_{2k+1,i}=w_{2k+1}^\dagger A u_i,\ i=1,\ldots,2k$ 
are nonzero. 
These elements can be computed without additional matrix-vector products at the cost
of storing two additional vectors.
Let $r_j$ and $\hat{r}_j$ be the last residuals that were added to the bases as vectors $v_m, w_m$ at iteration $j$.
The next basis vectors $v_{2k+1}$ and $w_{2k+1}$ after restart
are proportional to $r_{j+1}$ and $\hat{r}_{j+1}$. 
}
Thus, to compute the elements 
$T_{i,2k+1}$ and $T_{2k+1,i}$ it is sufficient to have $A r_{j+1}$ and $A^\dagger \hat{r}_{j+1}$.
To avoid additional matrix-vector multiplications we use the relations:
\begin{equation}
\begin{array}{rcl}
A r_{j+1} &=& Ap_{j+1} - \beta_j A p_j, \\
A^\dagger \hat{r}_{j+1} &=& A^\dagger \hat{p}_{j+1} - \bar{\beta}_j A^\dagger \hat{p}_j. 
\end{array}
\label{eq:TheApVecs}
\end{equation}
The vectors $Ap_{j+1}$ and $A^\dagger \hat{p}_{j+1}$ are available at iteration $j$ in \bicg, while
the vectors $Ap_j$ and $A^\dagger \hat{p}_j$ are specifically stored in \eigbicg. Note that copying the vectors 
$Ap_j$ and $A^\dagger \hat{p}_j$ to their storage is only needed just before restarting and not
in every iteration. 
Starting from the $(2k+2)$-{th} vectors, the elements
of the projection matrix are given by the three-term recurrence in equations (\ref{eq:tridiagonal}). The structure of the
projection matrix after any restart is given by:
\begin{equation}
T=W^\dagger A V = 
\left (
\begin{array}{c c c c c c c c c}
\gamma_1  &            &            &                & \times  &         &        &        &       \\
           &  \gamma_2 &            &                & \times  &         &        &        &       \\
           &            &  \ddots    &                & \times  &         &        &        &       \\
           &            &            & \gamma_{2k} & \times  &         &        &        &       \\
  \times   &    \times  &    \times  &      \times    &    \times    &     \times   &        &        &       \\
           &            &            &                &    \times    &     \times   &    \times   &        &       \\
           &            &            &                &         &  \ddots & \ddots &  \ddots&       \\ 
\end{array}
\right).
\end{equation}

\subsection{Algorithm implementation}
\label{subsec:algorithm_implement}
Figure \ref{fig:eigbicg_algorithm} shows the \eigbicg algorithm as an extension to \bicg.
It solves $Ax=b$ while computing $k$ approximate eigentriplets of $A$. 
The maximum size of the eigenvalue search space is $m$.

\begin{figure}[htbp]
\begin{center}
\fbox{
\begin{minipage}[t]{\textwidth}
\begin{tabbing}
\=x\=xxxxxx\=xxx\=xxx\=xxx\=x\=x\=\kill
{\bf \eigbicg algorithm: solve $Ax=b$ and compute $k$ approximate eigenvalues}\\
\> \\
\> (0)   \>\> $r_0=b-Ax_0,\ p_0=r_0$\\
\>       \>\> Choose $\hat{r}_0$ such that $<\hat{r}_0,r_0> \neq 0$\\
\>       \>\> $\hat{p}_0=p_0$\\
\>\> (0.1) \> $\eta=Ap_0,\ \hat{\eta}=A^\dagger \hat{p}_0$\\
\>       \>\> $\rho_0=<\hat{r}_0,r_0>$, if $\rho_0=0$ stop\\
\>\> (0.2) \> $\theta_0=\frac{1}{\sqrt{|\rho_0|}}, \ \delta_0=\frac{1}{\theta_0 \bar{\rho}_0}$\\ 
\>       \>\> $\beta_{-1}=0, \ \tau_0=<\hat{p}_0,\eta>, \ \tau_{-1}=0, \ l=0$\\
\>\> (0.3) \> update\_ev $=$ true\\
\> (1)   \>\> for $i=0,1,..$ till convergence do \{ \\ 
\>\> (1.1) \> \> if (update\_ev) \{$l=l+1$, $v_l=\theta_i r_i, \ w_l=\delta_i\hat{r}_i$\}\\
\> (2)   \>\> \> $\alpha_i=\frac{\rho_i}{\tau_i}$\\
\>(3--5) \>\> \>  $x_{i+1} = x_i + \alpha_ip_i, r_{i+1} = r_i - \alpha_i\eta,\ \hat{r}_{i+1} = \hat{r}_i - \bar{\alpha}_i \hat{\eta}$\\
\> (6)   \>\> \> $\rho_{i+1}=<\hat{r}_{i+1},r_{i+1}>$, if $\rho_{i+1}=0$ stop\\
\>\>(6.1)  \> \> if (update\_ev) \{$\theta_{i+1}=\frac{1}{\sqrt{|\rho_{i+1}|}}$, $\delta_{i+1} =  \frac{1}{\theta_{i+1} \bar{\rho}_{i+1}}$\}\\
\>(7--9) \>\> \> $\beta_i = \frac{\rho_{i+1}}{\rho_{i}}$, 
                    $p_{i+1}=r_{i+1} +\beta_i p_i, \ \hat{p}_{i+1}= \hat{r}_{i+1} + \bar{\beta}_i \hat{p}_i$\\
\>\>(9.1)  \> \> if (($l=m$) \& (update\_ev)) \{$\xi=\eta, \ \hat{\xi}=\hat{\eta}$\}\\
\>\>(9.2)  \> \> $\eta=Ap_{i+1}, \ \hat{\eta}=A^\dagger \hat{p}_{i+1}$, $\tau_{i+1}=<\hat{p}_{i+1},\eta>$\\ 
\>\>(9.3)  \> \> if (update\_ev) \{ \\
\>\>(9.4)  \> \> \> $T_{l,l}=\bar{\delta}_i \theta_i (\tau_i + \beta_{i-1}^2 \tau_{i-1})$\\
\>\>(9.5)  \> \> \> if ($l<m$) \{$T_{l,l+1}=-\bar{\delta}_i \theta_{i+1} \beta_i \tau_i$, 
                           $T_{l+1,l}=-\bar{\delta}_{i+1} \theta_i \beta_i \tau_i $\}\\
\>\>(9.6)  \> \> \> if ($l=m$) \{ \\
\>\>(9.7)  \> \> \>\> if ($w_m^\dagger V^{(m-1)}> (m-1) btol$) update\_ev $=$ false\\  
\>\>(9.8)  \> \> \>\>\> ($btol$ is a tolerance for biorthogonality loss (see section \ref{subsec:effect_loss_biorth}))\\
\>\>(9.9)  \>  \>\>\> Use the algorithm in Figure \ref{fig:bicg_eigpart} to compute Ritz triplets \\
\>\>(9.10)  \>  \>\>\> using $\Vm$, $\Wm$ and $T=\Wmdagger A \Vm$ and restart \\
\>\>(9.11)  \>  \>\>\> Set $T_{2k+1,j}=\bar{\delta}_{i+1}<\hat{\eta}-\bar{\beta}_i \hat{\xi},v_j>$, for $j=1,2,\ldots,2k$\\
\>\>(9.12)  \>  \>\>\> Set $T_{j,2k+1}=\theta_{i+1}<w_j,\eta-\beta_i \xi>$, for $j=1,2,\dots,2k$\\
\>\>(9.13)  \>  \>\>\> Set $l=2k$\\
\>\>       \> \> \> \}\\
\>\>       \> \> \}\\                           
\>\>       \> \} \\
\>\>       \> Compute final eigenvectors and eigenvalues before returning:\\
\>\> (10.1)\>({\it optional}) Biorthogonalize $V^{(l)}, W^{(l)}$ and recompute $T=W^{(l)^\dagger} A V^{(l)}$\\
\>\> (10.2)\> Compute the $k$ eigenvalues $\gamma_j$, right eigenvectors $f_j$,\\
\>\>       \>\> and left eigenvectors $g_j$ of interest of $T,\ j=1,2,..,k$\\
\>\> (10.3)\> Return the $k$ Ritz values $\gamma_j$, right Ritz vectors $u_j$, and left Ritz vectors $q_j$\\
\>\>       \>\> where $u_j=V^{(l)}f_j$, and $q_j=W^{(l)}g_j,\ j=1,2,..,k$\\
\end{tabbing}
\end{minipage}
}
\caption{The \eigbicg algorithm. Steps that are whole digit numbers correspond to \bicg.}
\label{fig:eigbicg_algorithm}
\end{center}
\end{figure}

In terms of memory cost, the algorithm requires storage for
the six vectors normally stored in \bicg, i.e., $r_j$, $\hat{r}_j$, $p_j$, $\hat{p}_j$,  $A p_j$, $A^\dagger \hat{p}_j$.
In addition, the algorithm requires storage of $2m$ vectors for $\Vm$ and $\Wm$, two vectors $\xi$ and $\hat{\xi}$ 
for storing $Ap_j$ and $A^\dagger \hat{p}_j$ in (\ref{eq:TheApVecs}), plus small matrices of order $m$.
So, the additional storage cost in comparison to \bicg is $O((2m+2)n+m^2)$.

Computationally, the additional expense of \eigbicg over \bicg is the computation of the $2k$ left and right Ritz 
vectors at every restart  and the computation of the $4k$ 
elements $T_{i,2k+1}$ and $T_{2k+1,i}$, $i=1,2,\dots,2k$, using (\ref{eq:TheApVecs}). This amounts to $O(8k(m+1)n)$ flops at 
every restart. The flop count is less (20\% less) than a similarly restarted Arnoldi method: both methods restart a basis,
and while Arnoldi orthogonalizes new vectors at every iteration, \eigbicg restarts both left and right bases (see \cite{Nearly_Optimal_I} for a related complexity analysis).
The expense of solving small eigenvalue problems and biorthogonalizing vectors of size $O(m)$ is negligible. 

Before returning, \eigbicg computes the final $k$ eigenvalues and eigenvectors (steps (10.1--10.3)).
If solving for a single right-hand side, it is advisable to biorthogonalize the final set of basis vectors 
and recompute the projection matrix (step (10.1)) to guard against biorthogonality loss during the \bicg iterations.
The associated cost is $O(m^2)$ dot products and $O(m)$ matrix-vector multiplications.
If solving for multiple right-hand sides, we can simply compute the final $k$ eigenvectors based on the current bases 
since these will be biorthogonalized in the outer \increigbicg method (described in the following section).
Even then, step (10.1) might be advisable when a large degree of loss of biorthogonality is expected.

\subsection{Effect of loss of biorthogonality}
\label{subsec:effect_loss_biorth}
As in the symmetric \lanczos method, the nonsymmetric \lanczos vectors lose biorthogonality when 
Ritz values start to converge \cite{Tong00_errorBICG,Bai_94}. In addition, biorthogonality is lost 
due to round off in near-breakdown situations.
In this paper we assume that no breakdown occurs. For look-ahead techniques to avoid near-breakdowns 
we refer the reader to \cite{Taylor:1982,Parlett:1985,Freund:1990:p1,Freund:1990:p2}. 
Loss of orthogonality or biorthogonality in linear systems is less of a problem since 
it leads to the \lanczos method taking more iterations to converge. For eigenvalue
problems, loss of orthogonality has more serious effects: it leads to spurious eigenvalues and eigenvectors, 
limits the attainable accuracy of computed eigenvalues, and if left unchecked could reduce the achieved 
accuracy of already converged eigenvalues.

One solution is to apply selective biorthogonalization of the \bicg residuals with 
respect to the almost converged Ritz vectors in $\Vm$ and $\Wm$.
To avoid this significant expense, we opt instead to stop updating the
Ritz vectors when the monitored loss of biorthogonality of $\Vm$ and $\Wm$ reaches a 
user-specified threshold. Instead of an expensive check with $\|I - \Wmdagger \Vm\|$, we monitor
the biorthogonality loss of the last vector before restart, $w_m$.
If $w_m^\dagger V^{(m-1)}> (m-1) btol$, we stop updating $\Vm$ and $\Wm$
and let \bicg converge to the linear system.
Although this check occurs only at every restart, we can further reduce its expense if we only start
monitoring it after some Ritz vectors have sufficiently converged. 
The residual norm of the $k$-th Ritz vector is given by the well known formula:
$|T_{k+1,k} z_{kr} v_{k+1}|$, and thus can be monitored at no additional expense.

\section{Systems with multiple right-hand sides}
\label{sec:incr-eigbicg}
In this section, we describe the \increigbicg algorithm for solving multiple right-hand sides. 
The algorithm uses an outer basis to accumulate and improve eigenvectors found by subsequent runs of \eigbicg 
and uses deflation to accelerate convergence.

\subsection{Deflating \bicg and \bicgstab}
\label{sec:init-bicgstab}
Let $U_r^{(k)}$ and $U_l^{(k)}$ be two $n\times k$ matrices whose columns are 
approximate right and left eigenvectors of $A$ such that
$U_l^{(k)\dagger} U_r^{(k)}=I$.
There are several
ways to deflate \bicg or \bicgstab for solving a linear system of equations.
One popular way is to use an explicitly deflated operator $A$ by applying 
a projector at each iteration.
Similarly, one can use a spectral preconditioner for $A$.
This way, the Krylov method finds solutions in the complement of 
$U_r^{(k)},U_l^{(k)}$ 
\cite{Frank:2001,Luscher:2007se,Morgan:2002,Wang_Sturler_Paulino}. 
By projecting at every Krylov iteration this approach 
guarantees that no directions in $U_r^{(k)}, U_l^{(k)}$ are repeated 
and thus achieves the most effective deflation.
However, for the same reason, it can become prohibitively expensive with 
large deflation subspaces.
In \cite{eigPCG} we advocated that the simpler option of deflating the initial guess can be made to work 
equally well.
Let $x_0$ be a given initial guess of the linear system $Ax=b$.
A deflated initial guess will be given by
\begin{equation}
\tilde{x}_0=x_0 + U_r^{(k)} (U_l^{(k)\dagger} A U_r^{(k)})^{-1} U_l^{(k)\dagger}(b-Ax_0).
\label{eq:initbicg}
\end{equation}
This approach is called \initbicg and \initbicgstab (as an extension of the symmetric {\em init-CG} \cite{initCG}). 
When $U_r^{(k)}$ and $U_l^{(k)}$ are exact eigenvectors, and in exact arithmetic, \initbicg and \initbicgstab should converge as fast as if 
$U_r^{(k)}, U_l^{(k)}$ were projected at every step.
However, when these vectors are accurate only to a certain tolerance,
deflation in \initbicg and \initbicgstab will be effective only till the linear system converges
roughly to the same tolerance.  After that point, convergence will be similar to undeflated \bicg and \bicgstab. 
We avoid this problem by restarting \initbicg and \initbicgstab when this tolerance is reached.
The restarted residual is deflated again using (\ref{eq:initbicg}), and therefore the linear system converges 
with deflated speed until the same relative tolerance is achieved again.
In \cite{eigPCG} we found that 1--2 restarts are sufficient for \cg
to achieve convergence similar to a fully projected system with exact eigenvectors.

\subsection{Incrementally increasing eigenvector accuracy and number}
After solving a single linear system using \eigbicg, the number and accuracy of the computed eigenvalues is 
not sufficient to effectively deflate \bicgstab for subsequent systems. 
This is because when the linear system converges, typically only the smallest eigenvalue
  is computed to a similar accuracy while the rest of the eigenvalues that are necessary 
  for deflation have lower accuracy.
In addition, the limited search space in \eigbicg can only hold information for a small number $k$ of eigenvalues.
One could run the \eigbicg further until all required eigenvectors are obtained. However, this would be
similar to applying an eigensolver as a preprocessing phase. 
Instead, we extend the method we developed for the symmetric case to improve the number and accuracy of 
the computed eigenvalues while solving linear systems. We divide the method into two phases. 

In the first phase, we solve a subset $n_1$ of the systems using \eigbicg. 
With each linear system solved, a new set of left and right Ritz vectors $Q_l$ and $Q_r$ are computed with \eigbicg. 
These new vectors are biorthogonalized and appended to the current deflation subspaces, $U_l$ and $U_r$.
These incrementally built spaces are then used to deflate the next right-hand side 
using (\ref{eq:initbicg}). This deflation not only speeds up the next linear system but also guarantees that
\eigbicg will produce Ritz vectors in the complement of the previous $Q_l$ and $Q_r$.

At the end of the first phase, we have accumulated biorthogonal deflation subspaces $U_l$ and $U_r$ 
of dimension $n_1 k$. 
In the second phase, we use $U_l$ and $U_r$ to deflate \bicgstab for the next linear systems, $n_1+1,..,n_s$.
Since the eigenvectors computed in the first phase are not exact, \initbicgstab may need to be restarted
  as discussed in Section~\ref{sec:init-bicgstab}.

The resulting algorithm, \increigbicg, is described in Figure \ref{fig:increigbicg-algorithm}
and applies to systems with $n_s$ multiple right-hand sides for a non-defective matrix $A$.
The user specifies the number $n_1$ of right-hand sides that will be solved with \eigbicg.
This choice depends on computational and storage cost of the projector.
$m$ and $k$ are the sizes of the search subspaces and the number of eigenvectors computed with \eigbicg,
and $tol$ is the tolerance to which the linear systems are solved. 
We restart \bicgstab when the linear system converges below the user specified $rtol$. 
This restarting tolerance is usually close to the accuracy of the computed eigenvalues.

Computationally, every call to \eigbicg in the first phase is followed by
  a biorthogonalization of the $k$ newly computed eigenvectors, 
  which costs $k(2s+k-1)$ axpy-dot operations when using (\ref{eq:initbicg}),
  where $s$ is the number of vectors in $U_l$.
In addition, to augment the projection matrix $H$ the algorithm costs $2k$ matrix-vector products and $sk$ dot products.
In the second phase the deflation projection is the only overhead, which is small given that few restarts 
of \bicgstab are used.

The algorithm as given in Figure \ref{fig:increigbicg-algorithm} requires the storage of $2kn_1$ vectors in 
$U_l$ and $U_r$. Additionally, a temporary storage of $2m$ vectors is used by \eigbicg to compute $k$
approximate eigenvectors. 
Normally, storage of $2kn_1+2m$ vectors 
is not a problem as this number is on the order of the number of right-hand sides to be solved. Finally, $U_l$ and
$U_r$ are not used in \eigbicg or \bicgstab and can be kept in a secondary storage.

\begin{figure}[htbp]
\begin{center}
\fbox{
\begin{minipage}[t]{\textwidth}
\begin{tabbing}
xxxx\=xxx\=xxx\=xxx\=xxx\=\kill
{\bf \increigbicg algorithm for solving $Ax_i=b_i,\ i=1,2,..,n_s$}\\
\> \\
{\bf Input:}$\ m, k, tol, btol, rtol\geq tol, n_1 < n_s,$ and $x_{i0}$ initial guesses for $x_i$\\
{\bf Output:} Solutions $x_i$, deflation subspaces $U_l$, $U_r$, and $H=U_l^\dagger A U_r$\\
\>\\ 
\underline{First phase: Solve $n_1$ systems using \eigbicg.}\\
(1) \> for $i=1,2,\dots,n_1$ do \\
(2) \> \> if ($i=1$) $\ \tilde{x}_{i0}=x_{i0}\ $ else $\ \tilde{x}_{i0} =x_{i0}+ U_r H^{-1} (U_l^\dagger (b-Ax_{i0}))$\\
(3) \> \> Solve $Ax_i=b_i$ with $\tilde{x}_{i0}$ as initial guess to tolerance $tol$ using \eigbicg\\
    \> \> with search space of size $m$ and obtain $k$ biorthogonal eigenvectors \\
    \> \> $Q_l$ and $Q_r$\\
    \> \> if ($i=1$) \\
(4) \> \> \> $U_l=Q_l$, $U_r=Q_r$, and $H=U_l^\dagger A U_r$\\
    \> \> else \{ \\
(5)\>\>  \> Biorthogonalize ($Q_r,Q_l)$ against $(U_r,U_l)$ to get $(Q'_r,Q'_l)$\\
(6)\>\>  \> Extend the projection matrix:\\ 
     \>\>  \> \> $
             H = \begin{pmatrix}
                 H   &   U_l^\dagger A Q'_r \\
                 (Q'_l)^\dagger AU_r & (Q'_l)^\dagger A Q'_r \\
                \end{pmatrix} \nonumber
          $ \\
(7)\>\>  \> Append the new vectors to the deflation subspaces:\\
    \> \>  \> \>$U_l \leftarrow [U_l \quad Q'_l]$ and $U_r \leftarrow [U_r \quad Q'_r]$\\
    \> \>\} \\
\>\\
\underline{Second phase: Solve remaining systems with deflated restarted \bicgstab}\\
(1)\> for $i=n_1+1,\dots,n_s$ do \\
(2) \>\> $\delta = rtol$\\
(3) \>\> repeat \\
(4) \>\>\> Set $\tilde{x}_{i0} =x_{i0}+ U_r H^{-1} (U_l^\dagger (b-Ax_{i0}))$\\
(5) \>\>\> Solve $Ax_i=b_i$ with $\tilde{x}_{i0}$ as initial guess using \bicgstab to tolerance $\max(tol,\delta)$\\
(6) \>\>\> Set $\delta = \delta \cdot rtol,\ x_{i0}=x_i$\\
(7) \>\> until converged to tolerance $tol$\\
\end{tabbing}
\end{minipage}
}
\caption{\increigbicg algorithm}
\label{fig:increigbicg-algorithm}
\end{center}
\end{figure}

\section{Numerical Experiments}
\label{sec:results}
We test a MATLAB implementation of \eigbicg and \increigbicg with matrices 
  from various applications. 
All computations are performed in double precision on a Linux workstation with quad core Intel Xeon W3530 
processors at 2.80GHZ with 8MB cache and 6GB of memory.
The right-hand sides are random vectors generated using the function {\tt rand()} in MATLAB. 
\subsection{Test Matrices}
We use the following test matrices in our numerical experiments:
\begin{itemize}
\item{\it Discretized partial differential operator:}
The matrix used in this test corresponds to the five-point discretization of the operator
\begin{equation}
L(u) = -\frac{\partial^2 u}{\partial x \partial x}-\frac{\partial^2 u}{\partial y \partial y}
	+\beta(\frac{\partial u}{\partial x} + \frac{\partial u}{\partial y})
\label{eq:pde}
\end{equation}
on the unit square with homogeneous Dirichlet conditions on the boundary.
First order derivatives are discretized by central differences. 
The discretization grid size is $h=1/(l+1)$ which yields a matrix of size $n=l^2$.
The matrix, which we scale by $h^2$, 
is real, nonsymmetric with a positive definite symmetric part 
($\frac{A+A^\dagger}{2} > 0$). 
We use $\beta=1$ and $l=50$ which gives a matrix size $n=2,500$.
The matrix is generated using the SPARSKIT software \cite{SPARSKIT} 
and is labeled as $PD$ in our tests.
\item{\it Examples from Sparse Matrix Collection:}
We use two examples from the University of Florida Sparse Matrix Collection \cite{UF-SPARMAT}. 
The first is the matrix {\it light\_in\_tissue} describing light transport in soft tissue.
This matrix is complex nonsymmetric with size $n=29,282$. 
The second is the matrix $Orsreg\_1$ from oil reservoir simulation. It is real, nonsymmetric 
indefinite matrix of size $n=2,205$.
\item{\it Examples from Lattice QCD:}
Lattice QCD methods \cite{Rothe:2005,Gupta:1998} study the theory of the strong nuclear
force (Quantum Chromodynamcis or QCD) between quarks and gluons \cite{Muta:1987,Donoghue:1992} as defined on a 
discrete space-time grid. Lattice calculations require the solution of linear systems $A x_i=b_i$ for many right-hand sides
\cite{Gusken:1999te,Wilcox:noise:1999,Bali:2009}, where $A$ is a large, sparse, nonsymmetric matrix 
called the {\it Dirac operator}. The matrix $A$ depends on the quark mass parameter $m_q$ and the background gauge field.
In our tests we use Wilson discretization for quarks in which case the Dirac operator has the form
\begin{equation}
A = (m_q+4)I - \frac{1}{2} D,
\end{equation}
where I is a unit matrix and $D$ is a matrix that depends on the gauge field. In addition, we use an
{\it even-odd} preconditioner, which is equivalent to first coloring the sites of the lattice as even-odd
  and then solving the Schur complement only on the even sites:
\begin{equation}
((m_q+4)^2I_{ee} - \frac{1}{4} D_{eo} D_{oe}) x_e = (m_q+4)b_e + \frac{1}{2} D_{eo} b_o.
\label{eq:eo-precond}
\end{equation}
The subscripts {\it ee, eo, oe} refer to even-even, even-odd and odd-even lattice blocks 
respectively. Gauge fields were generated using the Wilson plaquette action and sea quark effects were ignored. We use 
two examples corresponding to the parameters given in Table \ref{Table:qcd-params}. The values of the mass parameter $m_q$ were
chosen such that quarks have very small mass in which case the system is nearly ill conditioned.
\begin{table}[htbp]
\caption{Parameters for the test QCD matrices}
\small{
\begin{center}
\begin{tabular}{|c|c|c|c|}
\hline
Matrix & Lattice Size & Gauge Coupling & $m_q$ \\
\hline
{\it QCD--49K} & $8\times8\times8\times8$  &  $5.5$ & $-1.25$ \\
{\it QCD-249K} & $12\times12\times12\times12$ & $5.8$ & $-0.95$ \\
\hline
\end{tabular}
\end{center}
}
\label{Table:qcd-params}
\end{table} 
\end{itemize}

\subsection{Stopping Criteria for linear systems}
In some of our numerical experiments, where we study the behavior of \eigbicg alone, 
we solve the linear system to a tolerance $tol$ which is close to machine double precision.
For these tests, we stop \eigbicg based on the criterion
$||r^{(i)}|| < tol(||A||_{est}*||x^{(i)}||+||b||)$, where $r^{(i)}, x^{(i)}$ are the \bicg residual and 
approximate solution at the $i$ step, and $||A||_{est}$ is an estimate of the norm 
of $A$ obtained inexpensively from the Lanczos iteration.
For our tests with \increigbicg we converge to higher tolerances $tol$ and therefore
we use the simpler criterion $||r^{(i)}|| < tol ||b||$.

\subsection{Benchmark algorithms}
The quality of the eigenvector approximations from \eigbicg depends on the
size of the search space and on how well it maintains biorthogonality against previous
\bicg residuals. To explore these effects, we compare the eigenvalues computed 
by \eigbicg with three benchmark algorithms:
\begin{itemize}
 \item Unrestarted \bilanczos: All the residuals generated while solving the linear system 
 are used to compute the approximate eigenspace. Comparing with this algorithm should 
 show the effect of using a small size subspace. However, loss of biorthogonality is present.
 \item Biorthogonalized \bilanczos: This is the same as {\it unrestarted} \bilanczos but with explicit
 biorthogonalization of the \bilanczos vectors. This should be the ideal algorithm since it is not
 affected by limited search space size or by loss of biorthogonality.
 \item \ortheigbilanczos: This is identical to \eigbicg 
with the exception that the \bicg vectors are explicitly biorthogonalized (twice) 
against all previously seen Lanczos vectors. In this case, only the limited subspace size
should have an effect on the computed eigenvalues.
\end{itemize}

\subsection{Results with \eigbicg}
We first demonstrate the properties of \eigbicg by exploring the following issues.
(1) the accuracy of the computed eigenvalues in comparison to the benchmark algorithms.
(2) the effect of biorthogonality loss on the computed eigenvalues.
(3) provide some guidance on choosing the subspace size, $m$, and the number of eigenvectors to compute, $k$.
\subsubsection{Comparing with benchmark algorithms.}
\begin{figure}[htbp]
\begin{center}
\includegraphics[width=0.47\textwidth ]{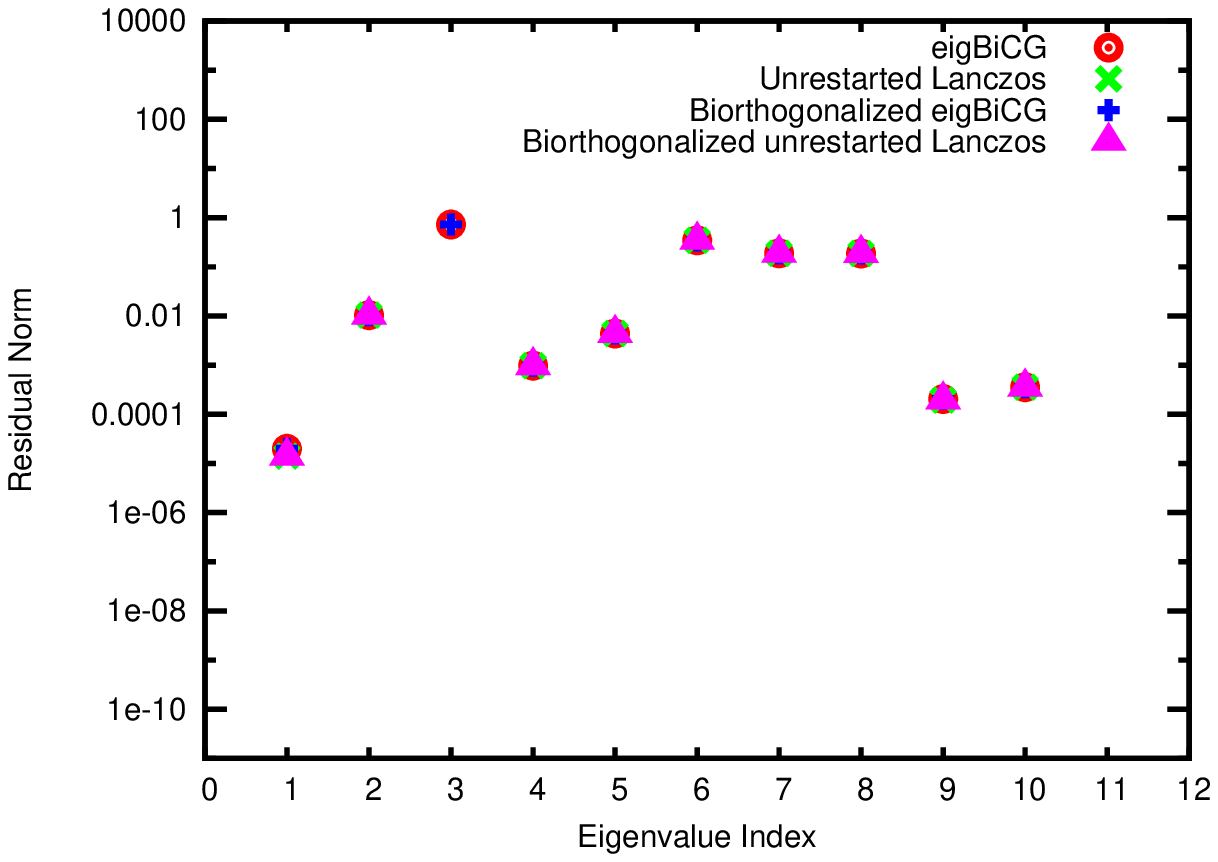}
\includegraphics[width=0.47\textwidth ]{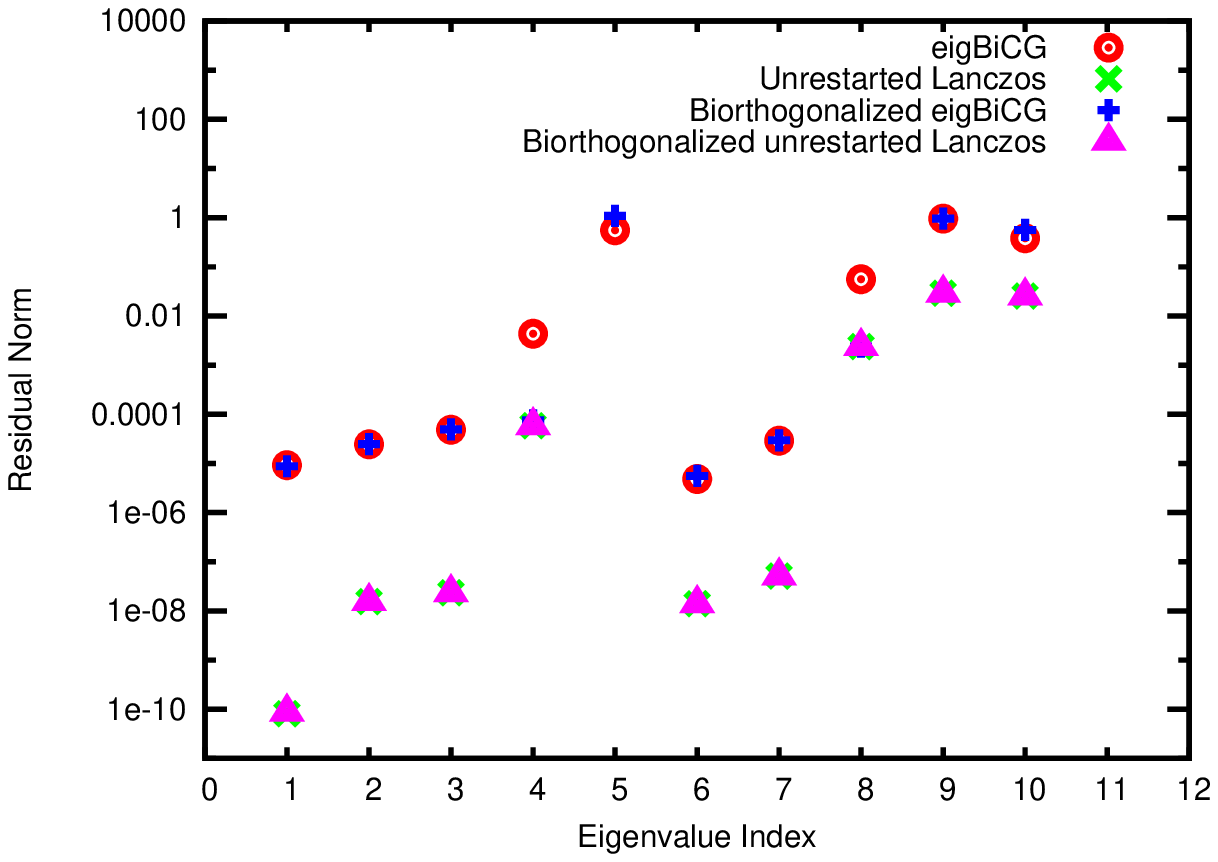}
\caption{Comparing eigenvalue residual norms obtained with \eigbicg and benchmark algorithms 
for the {\it QCD--49K} (left) and {\it QCD--249K} (right) matrices.}
\label{Fig:qcd-benchmark-compare}
\end{center}
\end{figure}

In the following tests, we solve the linear system to $tol=10^{-12}$ using
\eigbicg with $k=10, m=40$. Updating the eigenvectors stops after biorthogonality is lost to $btol = 10^{-4}$.
\begin{itemize}
 \item {\it $PD$ matrix:}
The linear system in this case converges in $172$ iterations. We observe that both \eigbicg and the
benchmark methods computed 10 Ritz values that were practically identical. Moreover, the norms of the 
residuals of the Ritz vectors, $||Aq-\lambda q||/||q||$, were all within $10^{-6}$ relative 
difference between methods. The only exception was the smallest eigenvalue, for which different methods
showed residual norms with $10^{-14}$ absolute difference.
Table \ref{Table:pdb1-2500} shows seven of the computed Ritz values and their residual norms 
(for only one method as they do not differ in the first 6 digits). Note that the smallest eigenvalue 
has converged to about the same accuracy as the the linear system.

\begin{table}[htbp]
\caption{Seven smallest Ritz value and their residual norms for PD matrix.}
\small{
\begin{center}
\begin{tabular}{llllllll}
\hline
RitzVal  & 7.78e-03 &1.91e-02& 3.05e-02& 3.80e-02& 4.94e-02& 6.44e-02& 6.83e-02\\
ResNorm & 1.11e-10 &3.40e-08& 3.98e-05& 1.97e-06& 1.21e-04& 2.57e-03& 4.03e-03 \\
\hline
\end{tabular}
\end{center}
}
\label{Table:pdb1-2500}
\end{table} 

\item {\it light\_in\_tissue matrix:}
In this case, the linear system converges in $436$ iterations. 
\cblue{All methods computed the same ten smallest eigenvalues with agreement in at least 6 relative
digits. Such good agreement is surprising given that \eigbicg used a subspace of size $m=40$, while unrestarted Lanczos
computed the same eigenvalues using a subspace of size $436$.}

\item {\it Orsreg1 matrix:}
This matrix is highly indefinite with several eigenvalues close to zero,
and all methods, including a fully biorthogonal Bi-Lanczos,
failed to approximate any eigenvalues. 

\item {\it QCD--49K matrix:}
The linear system in this case converges in $353$ iterations. \eigbicg found the same Ritz values 
as the other methods with at least 6 relative digits of accuracy, except for a single spurious 
eigenvalue. 
\cblue{The same (3rd smallest) spurious eigenvalue was produced also by the \ortheigbilanczos method, 
but not by the unrestarted \bilanczos. This implies that this is an artifact of the limited window size 
and not of the loss of biorthogonality.} In Figure \ref{Fig:qcd-benchmark-compare},
we show the residuals for the eigenvalues computed with different algorithms.

\item {\it QCD--249K matrix:} 
In this case, {\it eigBiCG(10,40)} converges to the linear system in $698$ iterations.
A similar behavior was observed as in the {\it QCD--49K} case. The six eigenvalues with smallest 
magnitude agree in 6 relative digits between all methods, 
while one spurious eigenvalue (the 5th) is produced by both \eigbicg and \ortheigbilanczos.
The 7th through the 10th eigenvalues had larger discrepancies. See Figure \ref{Fig:qcd-benchmark-compare}
for comparison of the eigenvalue residual norms computed by different methods.
\end{itemize}

\cblue{
The above observations, concurring with our experiments on several other matrices, suggest that
\eigbicg is able to compute approximations to a few smallest eigenvalues that are as accurate as
unrestarted \bilanczos, in spite of the limited size of the subspace used.
On the other hand, the limited size may cause an occasional spurious interior eigenvalue, 
  as evidenced by the fact that this appears only from \eigbicg and \ortheigbilanczos, 
  but not from unrestarted or biorthogonalized \bilanczos.
The failure of all benchmark algorithms on matrix {\it Orsreg1} shows the limitation of
the underlying \bicg method for indefinite matrices rather than \eigbicg.
}

\begin{figure}[htbp]
\begin{center}
\includegraphics[width=0.45\textwidth ]{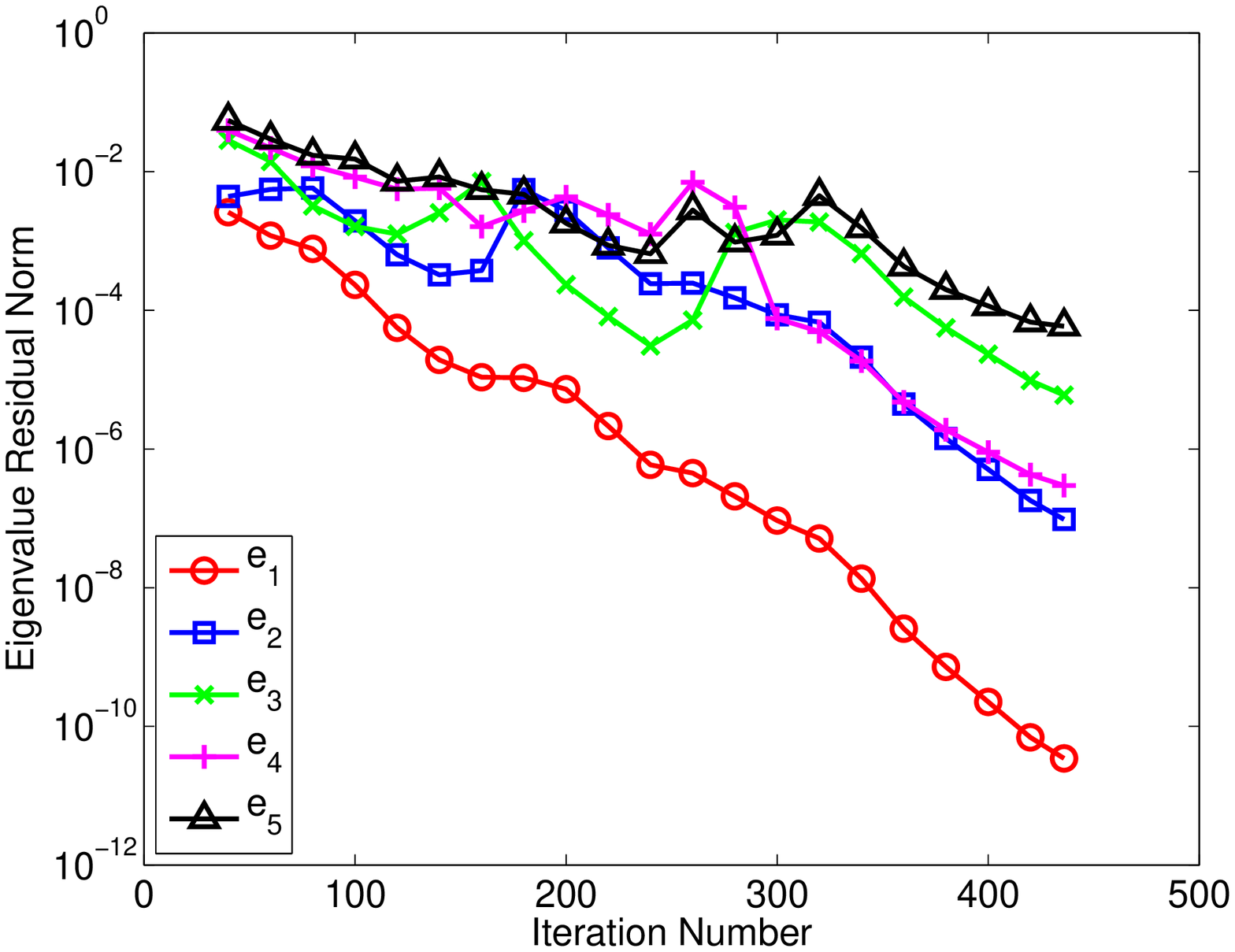}
\includegraphics[width=0.45\textwidth ]{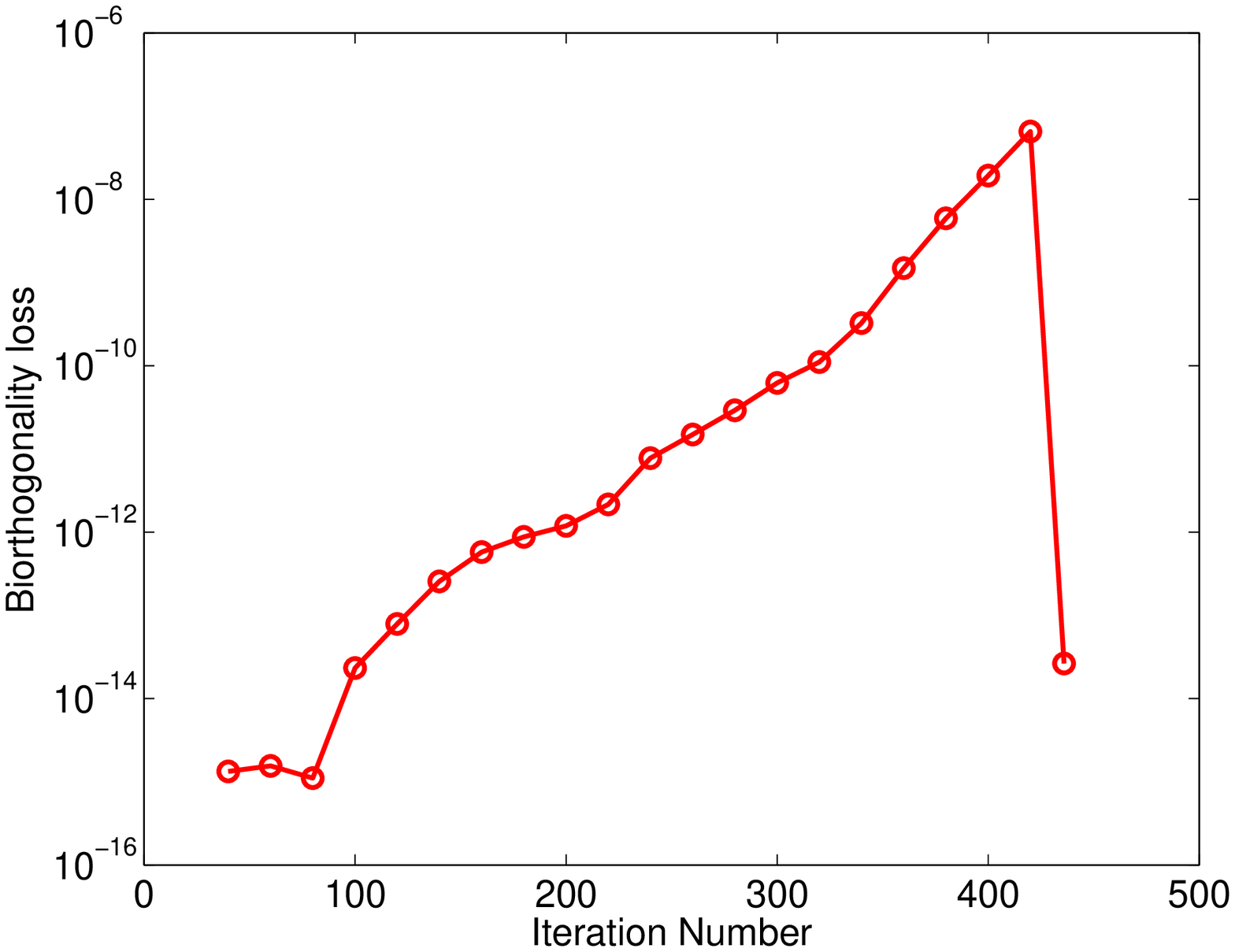}
\caption{{\it eigBiCG(10,40)} on the light\_in\_tissue matrix. Left: convergence history of the five smallest
eigenvalues. Right: Loss of biorthogonality between $\Vm, \Wm$.}
\label{Fig:light-ereshist}
\end{center}
\end{figure}

\cblue{
Figure \ref{Fig:light-ereshist} shows the convergence history of the \eigbicg for the
  five smallest eigenvalues of the matrix {\it light\_in\_tissue}.
Although not shown, the eigenvalue convergence history of the unrestarted \bilanczos is identical.
The right part of the figure plots $1-\|\Wmdagger \Vm\|$ as a measure of the loss of biorthogonality 
between left and right basis vectors.
As expected, this increases as the smallest eigenvalue converges.
}

\subsubsection{Choosing $m$ and $k$ for \eigbicg.} 
\begin{figure}[htbp]
\begin{center}
\includegraphics[width=0.35\textwidth ]{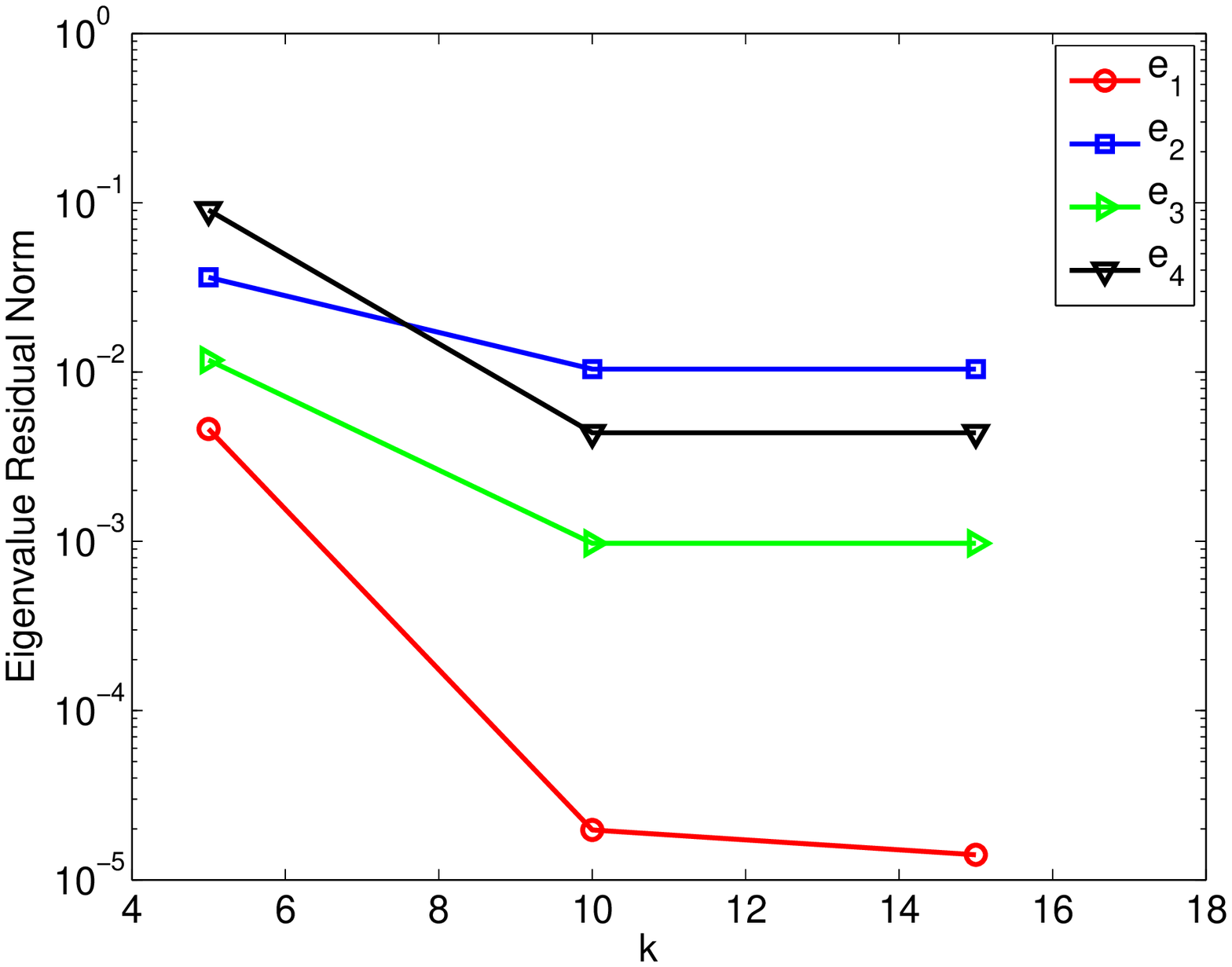}
\includegraphics[width=0.35\textwidth ]{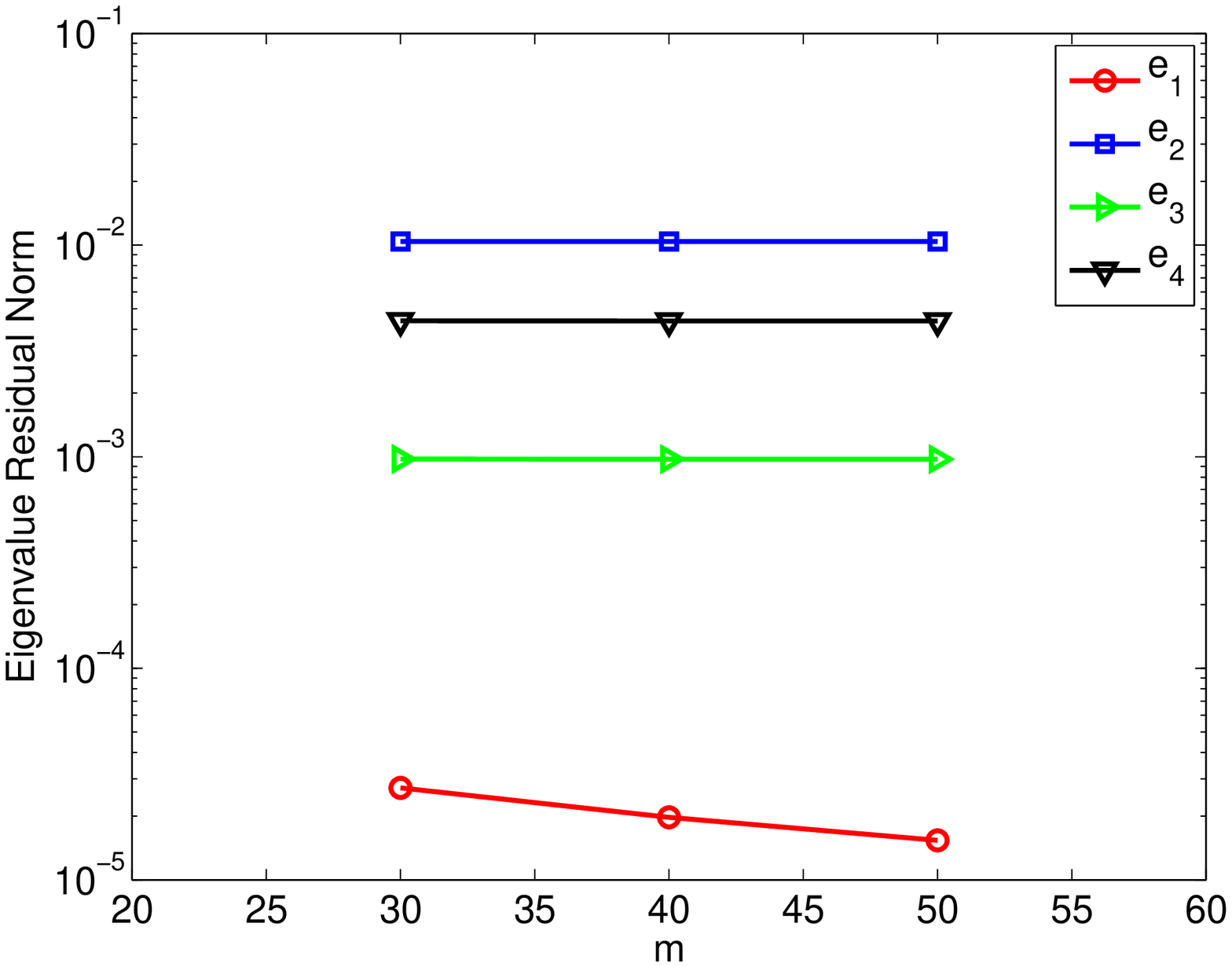}
\caption{QCD--49K matrix: Residual norms of 4 smallest eigenvalues.
Left: from {\it eigBiCG(k,40)} as a function of $k$. 
Right: from  {\it eigBiCG(10,m)} as a function of $m$.}
\label{Fig:qcd-49k-m-nev-depend}
\end{center}
\end{figure}

\begin{figure}[htbp]
\begin{center}
\includegraphics[width=0.35\textwidth ]{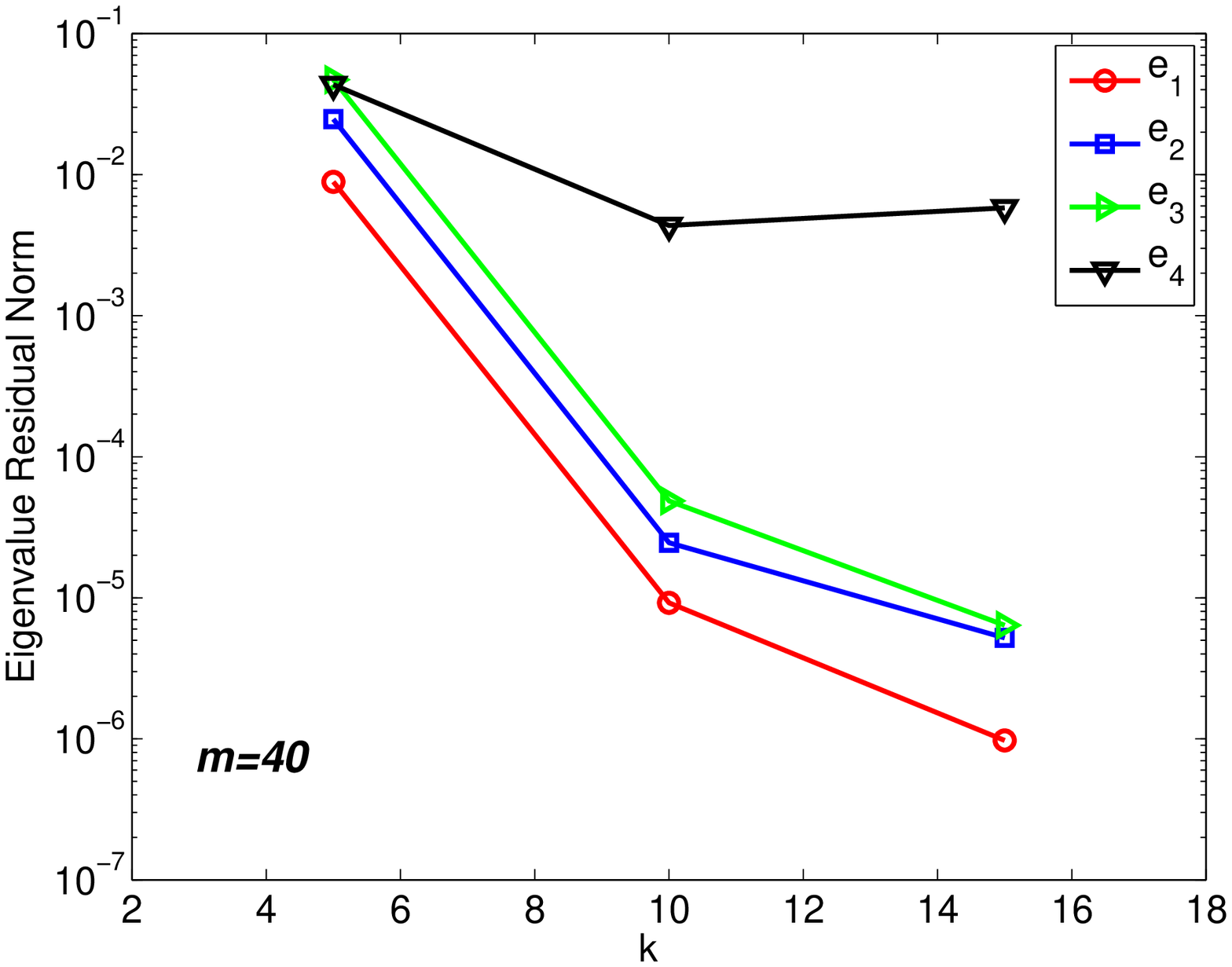}
\includegraphics[width=0.35\textwidth ]{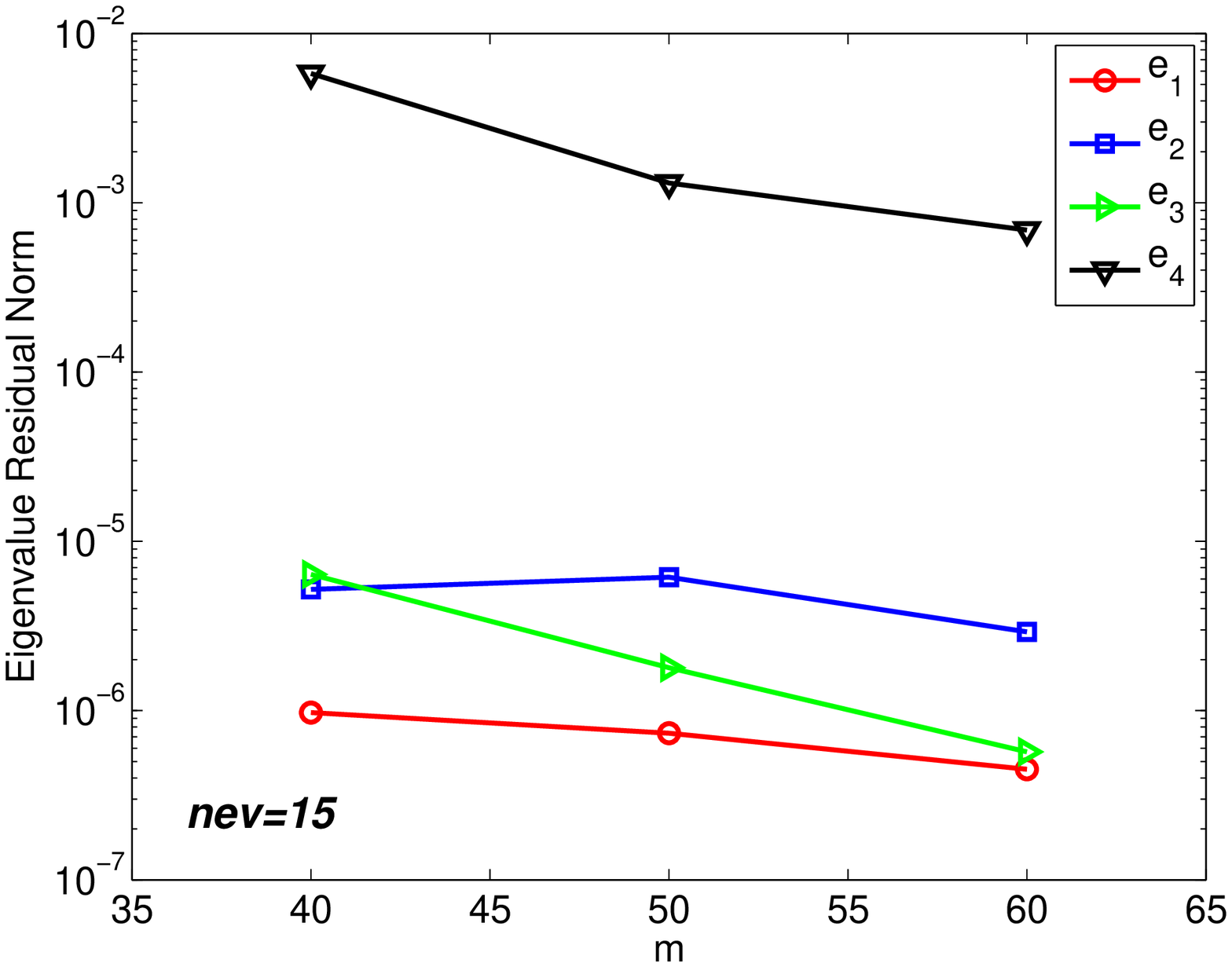}
\caption{QCD--249K matrix: Residual norms of 4 smallest eigenvalues.
Left: from {\it eigBiCG(k,40)} as a function of $k$. 
Right: from  {\it eigBiCG(15,m)} as a function of $m$.}
\label{Fig:qcd-249k-m-nev-depend}
\end{center}
\end{figure}

\cblue{
Beyond the condition $m > 2k$, the parameters $m, k$ should be chosen to minimize 
  the computational cost and approximate well as many eigenvalues as possible.
As we discussed earlier, \eigbicg is stopped when the linear system converges 
  so interior eigenvalues are not expected to be as accurate as the smallest one.
Therefore, choosing $k$ large in order to approximate more eigenvalues has 
  diminishing returns while increasing computational cost as $O(k^2)$.
On the other hand, the $2k$ vectors should encapsulate the information of 
  the whole $\Vm$ subspace at restart, so choosing $k$ too small deteriorates 
  eigenvalue convergence.
In our experiments we have observed that values of $k$ between 
  10 and 15 yield the best results.
Given a reasonable choice for $k$, we have observed that the accuracy of the eigenvectors 
  is not very sensitive to the value of $m$, so there is no reason to increase $m$ 
  too much. 
A typical choice such as $2k+10$ or $2k+20$ was found to be sufficient. 
An exploration of the effect of various choices of $m, k$ for the QCD matrices 
  is shown in Figures \ref{Fig:qcd-49k-m-nev-depend} and \ref{Fig:qcd-249k-m-nev-depend}.
These results are typical of other matrices as well.
A further fine-tuning of $m, k$ is also problem dependent, based on
  the conditioning of the matrix (as deflation benefits may be limited)
  and the number of right-hand sides.
}

\subsection{Experiments with \increigbicg}
We generated 21 random right-hand sides $b_i$.
The first 20 systems are solved using \eigbicg and the 21st system is solved using
{\it init-BiCGStab} that is deflated by the accumulated approximate eigenspace. The 21st system is also
solved using undeflated \bicgstab for comparison. 

\begin{figure}[htbp]
\begin{center}
\includegraphics[width=0.45\textwidth]{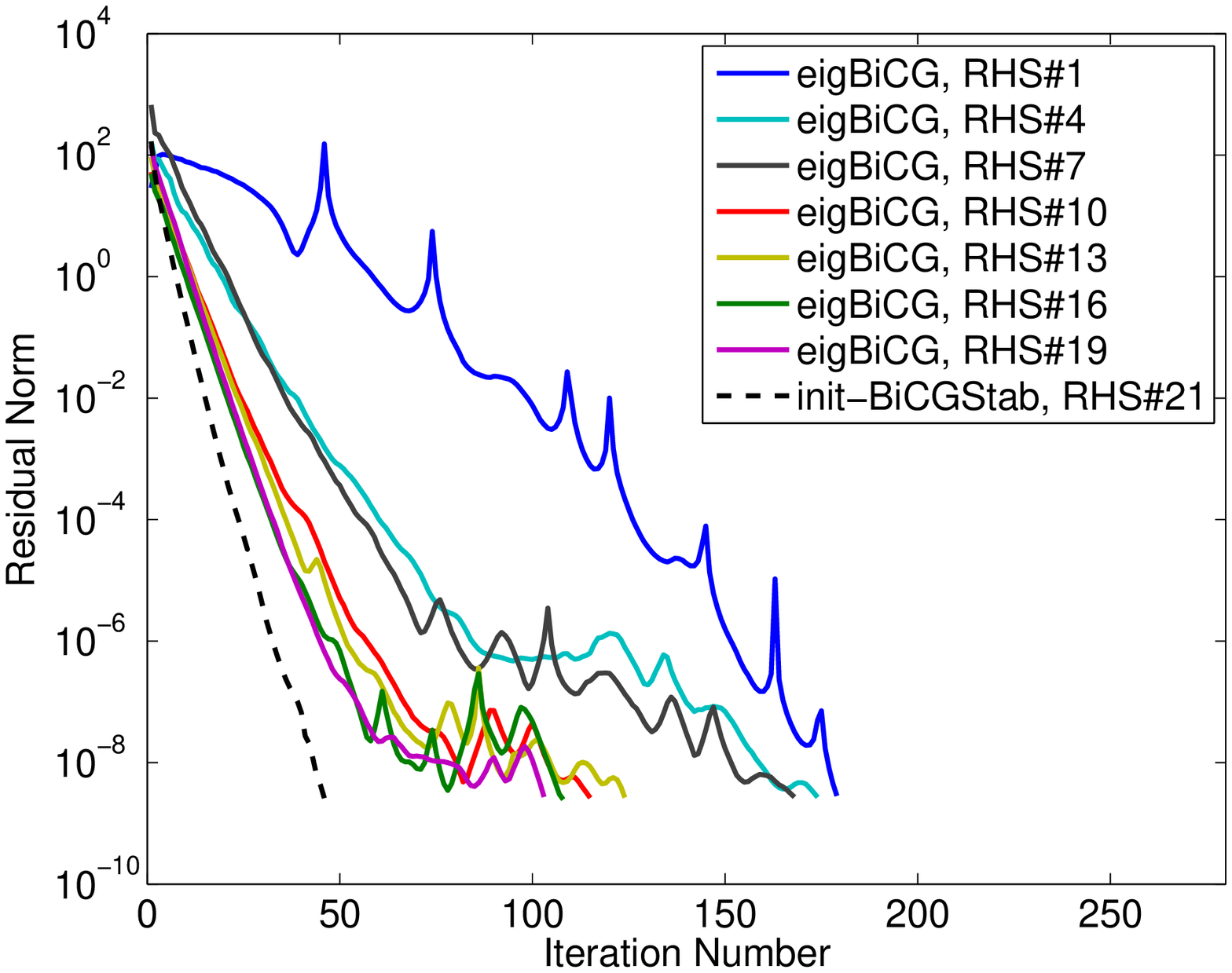}
\includegraphics[width=0.45\textwidth ]{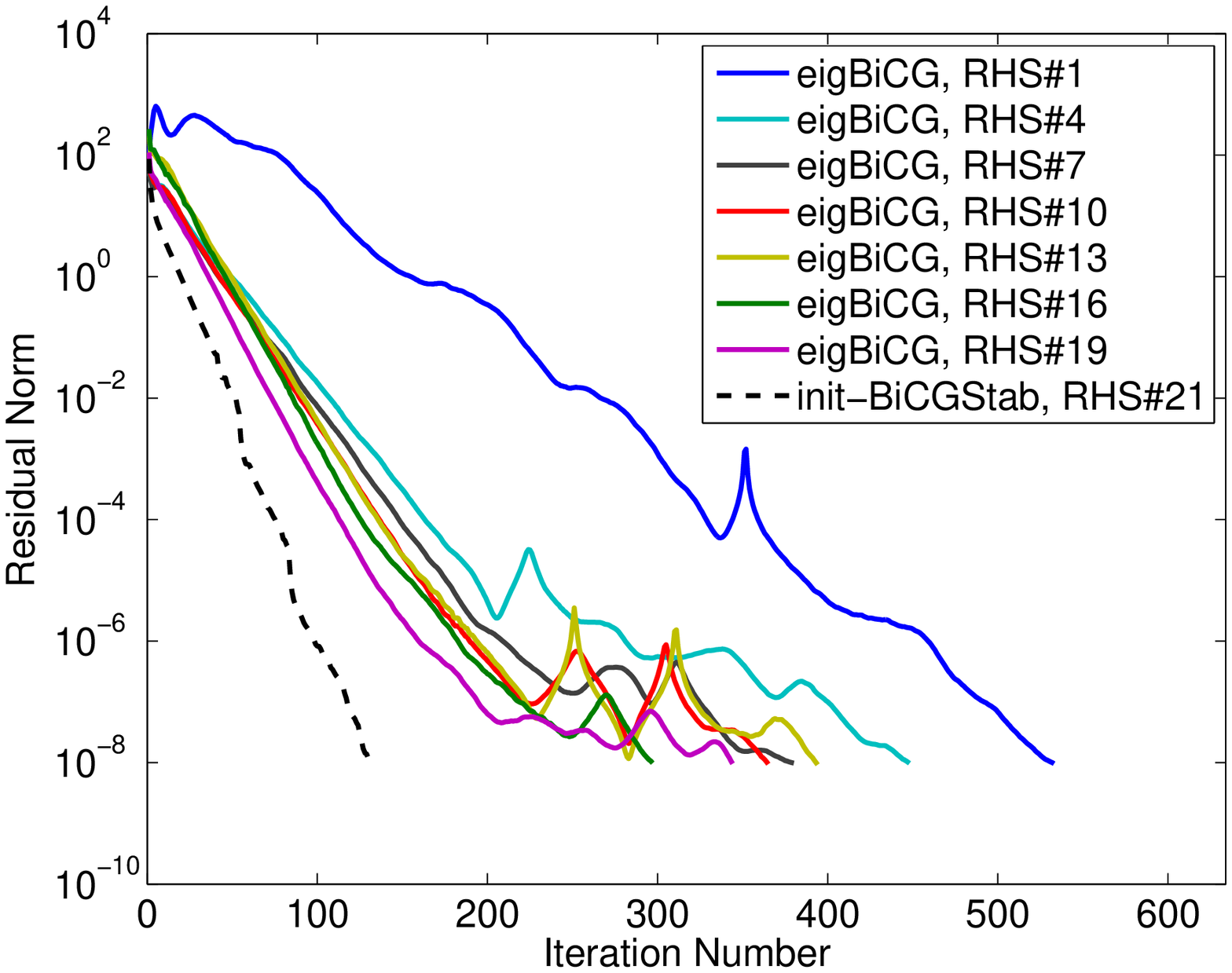}
\caption{
Convergence of some of the linear systems solved using \eigbicg and $init-$\bicgstab
for the matrix $PD$(left) and $light\_in\_tissue$(right). The first 20 systems are solved using \eigbicg(40,10),
and the 21st system is solved using \initbicgstab deflated with 200 eigenvectors.
}
\label{Fig:PD-and-Light-reshist}
\end{center}
\end{figure}

In Figure \ref{Fig:PD-and-Light-reshist}, we show the convergence of the 
residual norm
of every third linear system in phase one and for the 21st system (phase two)
for matrices $light\_in\_tissue$ and $PD$.
We use  $tol=10^{-10}$, $m=40$, $k=10$, 
and $btol=10^{-4}$.
We observe faster convergence as we solve more systems
  and deflate with more and better quality eigenvectors.
During the first phase, i.e. solving the first 20 systems 
using \eigbicg, the residual norm drops faster up to a certain value and 
then convergence slows down.
As we discussed earlier, when the linear system residual converges to a 
  tolerance comparable to the accuracy of the eigenvectors, the iteration 
  ``sees'' again the eigenvectors and deflation effects cease.
As more systems are solved, the eigenvectors improve incrementally,
  and thus the slow down occurs at lower tolerances.
If we restart and deflate again, we obtain faster convergence as we 
  see for the 21st system with \initbicgstab.

\begin{figure}[htbp]
\begin{center}
\includegraphics[width=0.447\textwidth]{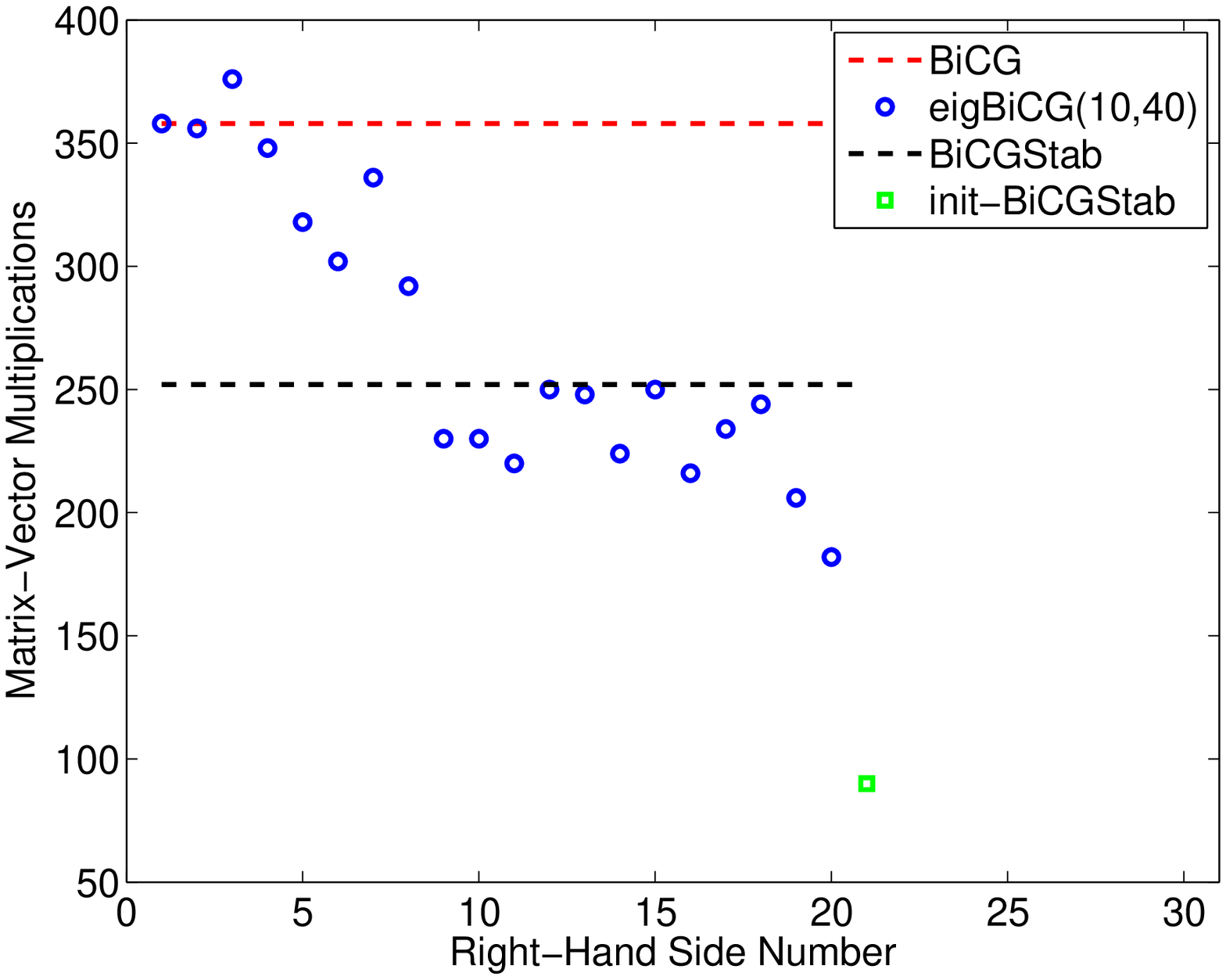}
\includegraphics[width=0.453\textwidth ]{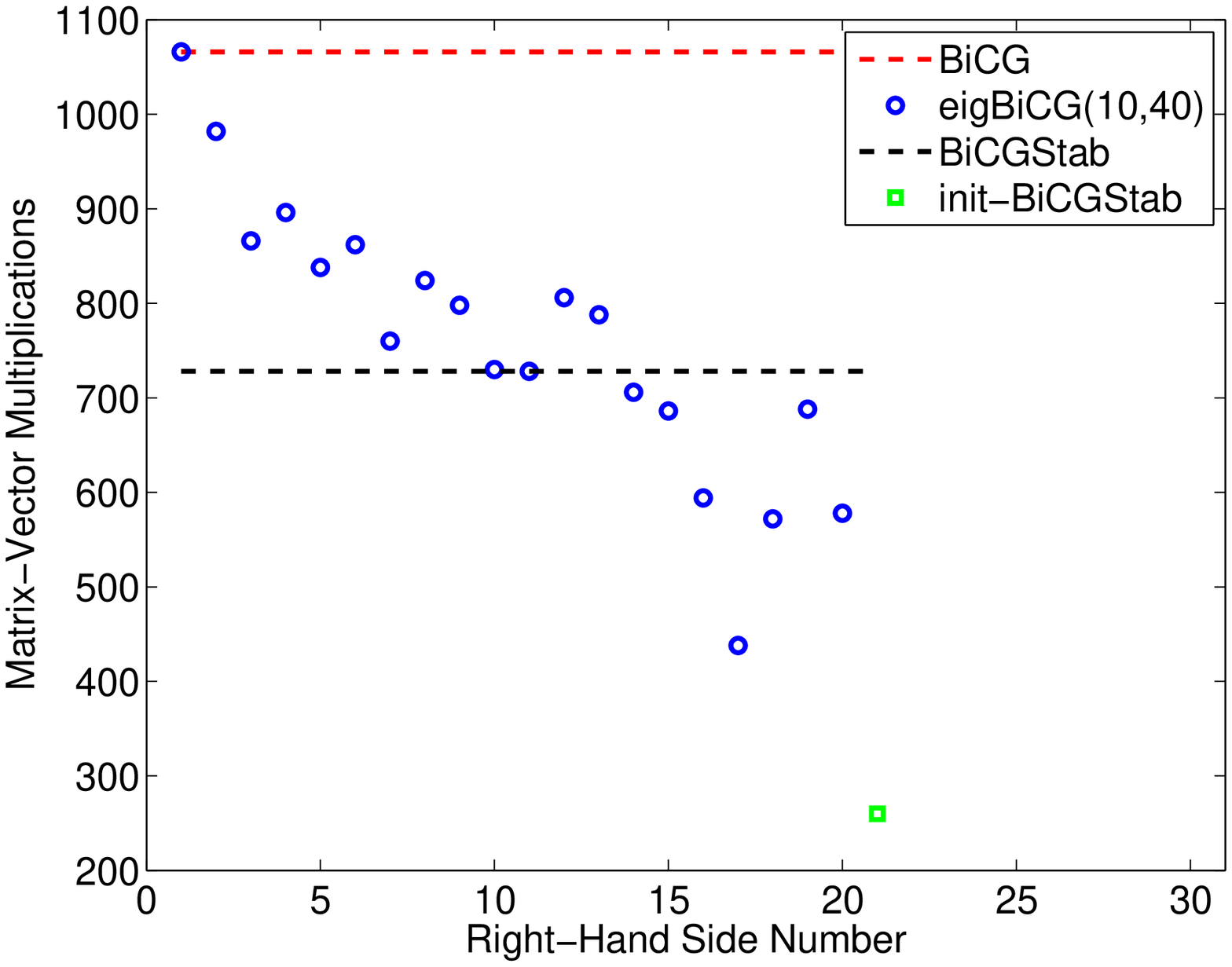}
\caption{
Reduction of the number of matrix-vector multiplications as we solve more systems for
the matrix $PD$(left) and $light\_in\_tissue$(right). For comparison, we also show the number of
matrix-vector multiplications using standard \bicg and \bicgstab. }
\label{Fig:PD-and-Light-matvecs-vs-rhs}
\end{center}
\end{figure}

In Figure \ref{Fig:PD-and-Light-matvecs-vs-rhs}, we show the number of matrix-vector multiplications
used to reach convergence for the 21 systems solved. We also show results for 
undeflated \bicg and \bicgstab, which are respectively 5 and 2.5 times slower than our method.

\begin{figure}[htbp]
\begin{center}
\includegraphics[width=0.45\textwidth]{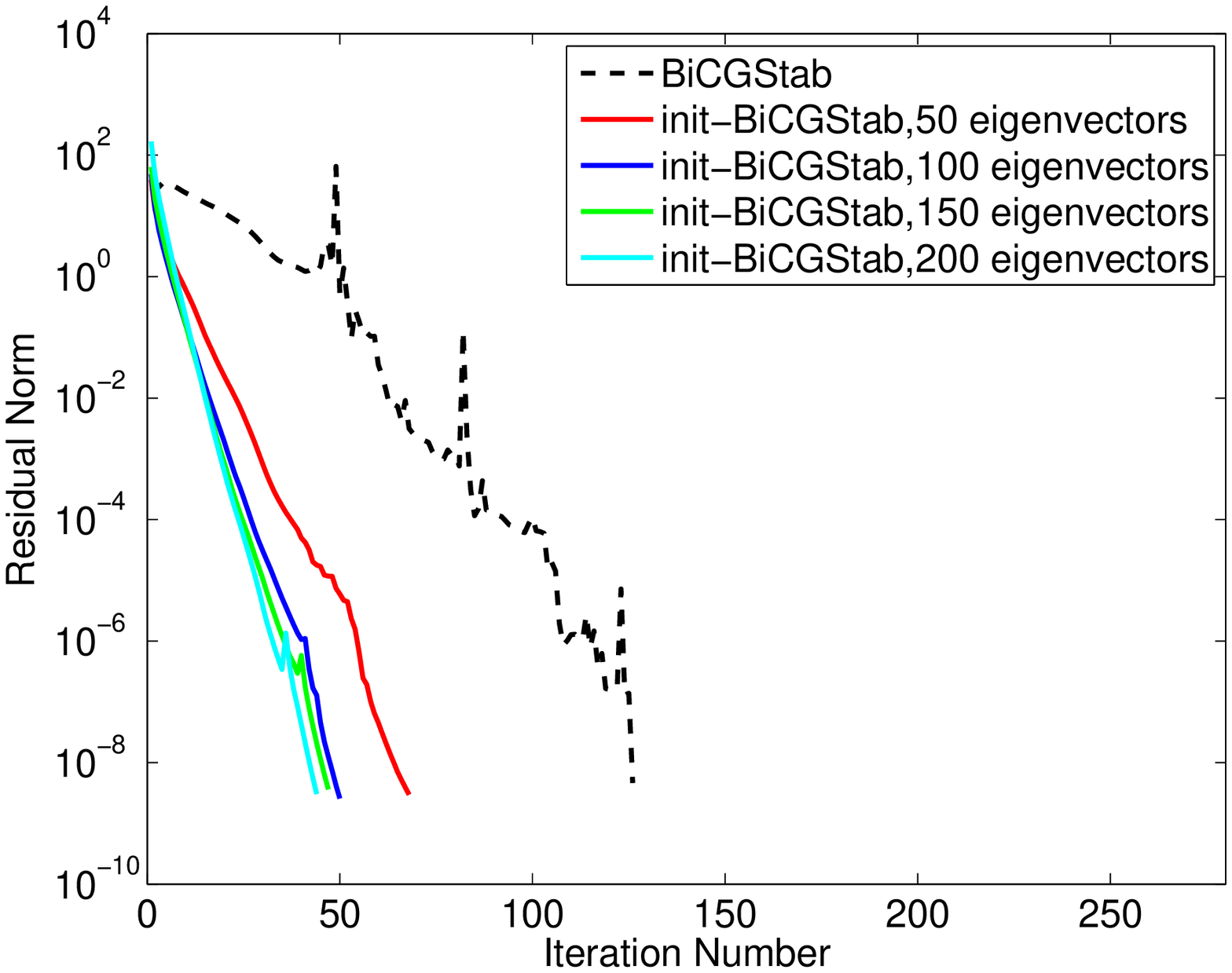}
\includegraphics[width=0.45\textwidth ]{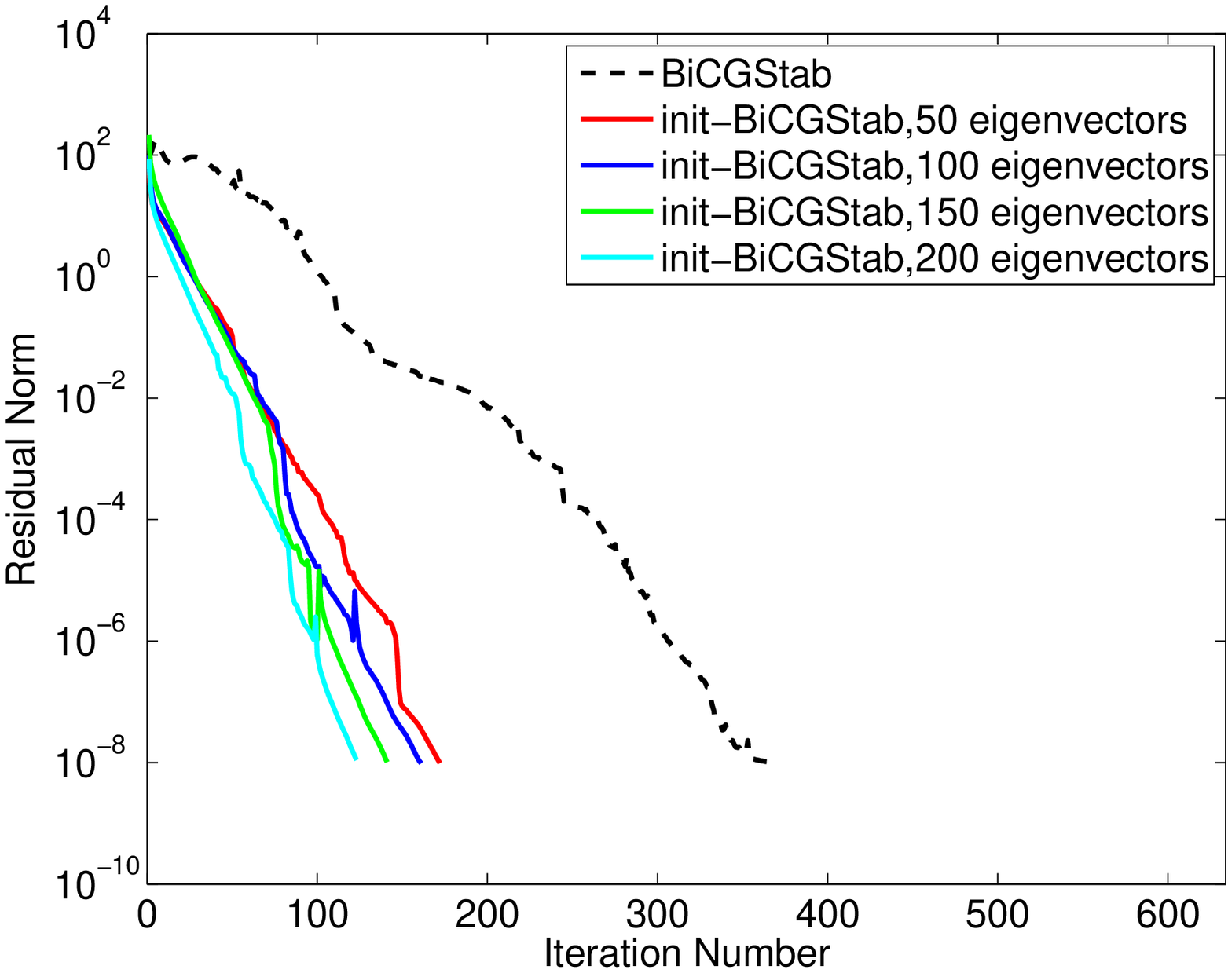}
\caption{
Effect of increasing the number of eigenvectors deflated on the number of iterations used by 
\initbicgstab for the matrices $PD$(left) and $light\_in\_tissue$(right). The plot shows that a small number 
of eigenvectors was enough to give the bulk of speedup.}
\label{Fig:PD-and-Light-dbstab-vs-nrhs}
\end{center}
\end{figure}

In Figure \ref{Fig:PD-and-Light-dbstab-vs-nrhs}, we compare the speedup obtained
for solving the 21st system with \initbicgstab when deflating with different 
numbers of approximate eigenvectors. For these problems, a modest
number of eigenvectors provide the most part of speedup. In general, 
this would depend on the distribution and clustering of the eigenvalues.
In the results shown above, \initbicgstab was restarted only once when the 
system converged to $rtol=10^{-8}$.

\begin{figure}[htbp]
\begin{center}
\includegraphics[width=0.453\textwidth]{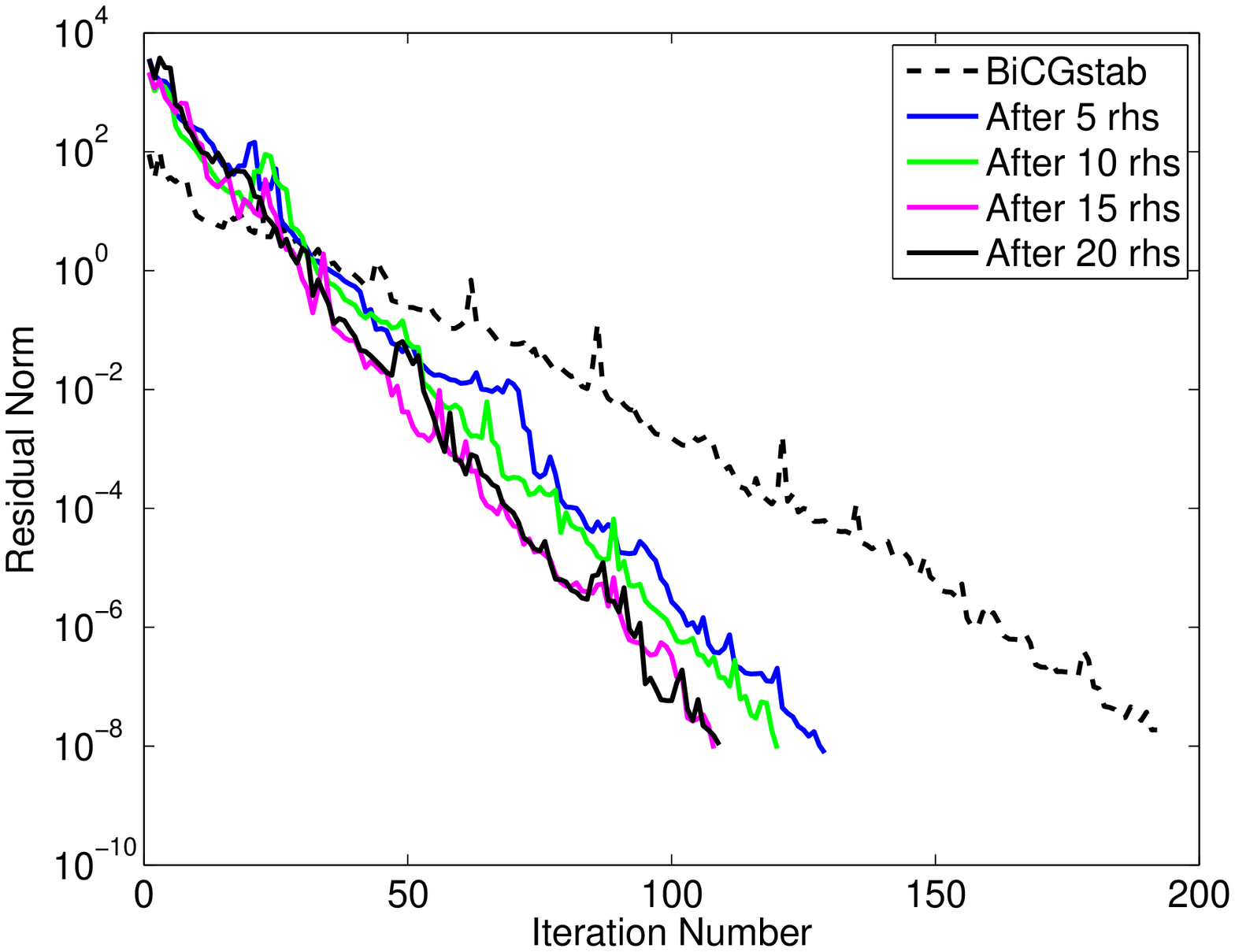}
\includegraphics[width=0.447\textwidth]{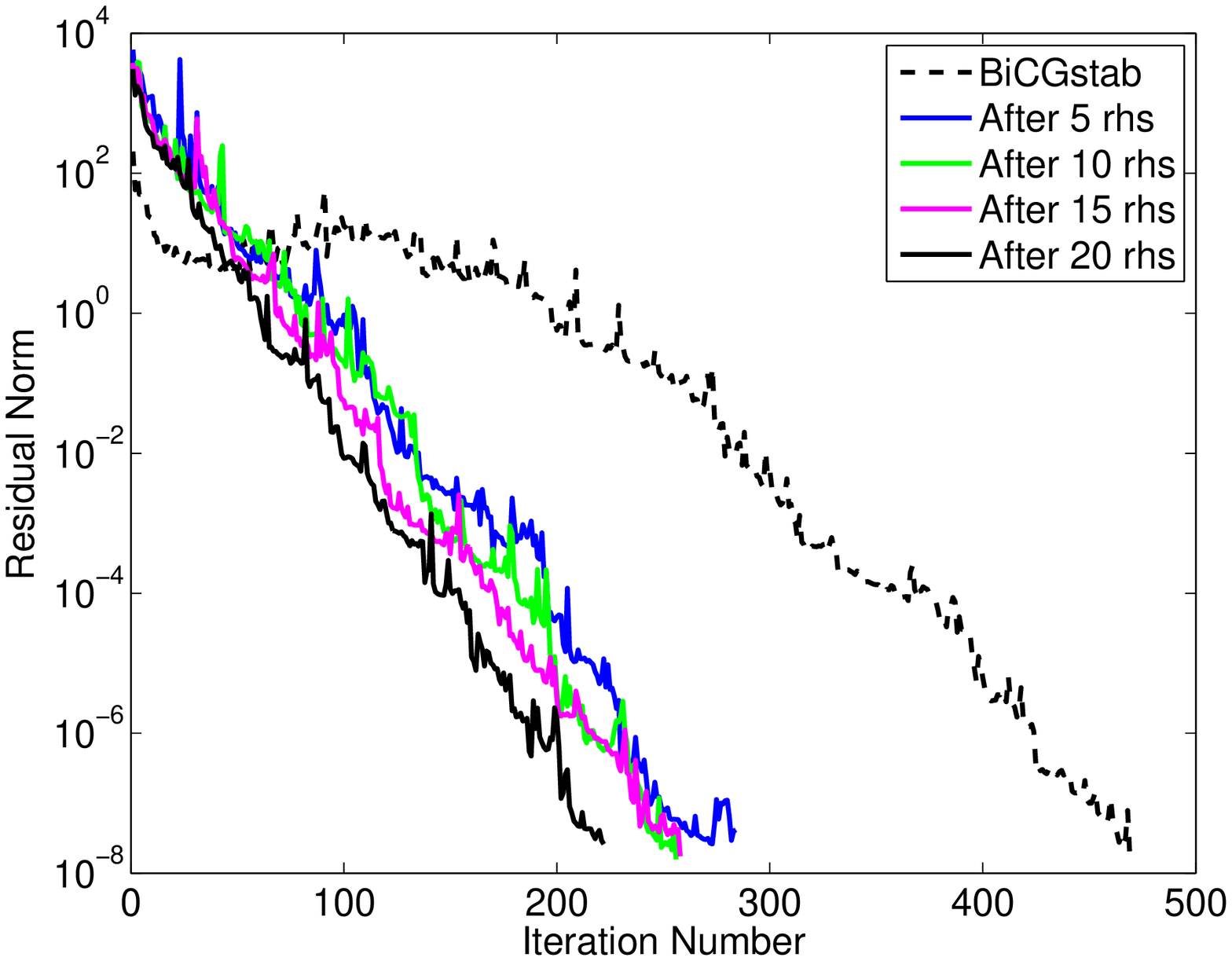}
\caption{
Convergence of undeflated \bicgstab versus $init-$\bicgstab deflated with the 
eigenvectors obtained after solving a different number of systems,
for the matrices $QCD-49K$(left) and $QCD-249K$(right). 
A small number of eigenvectors is enough to give most of the speedup.}
\label{Fig:qcd-defbstab-vs-nrhs}
\end{center}
\end{figure}

\begin{figure}[htbp]
\begin{center}
\includegraphics[width=0.45\textwidth ]{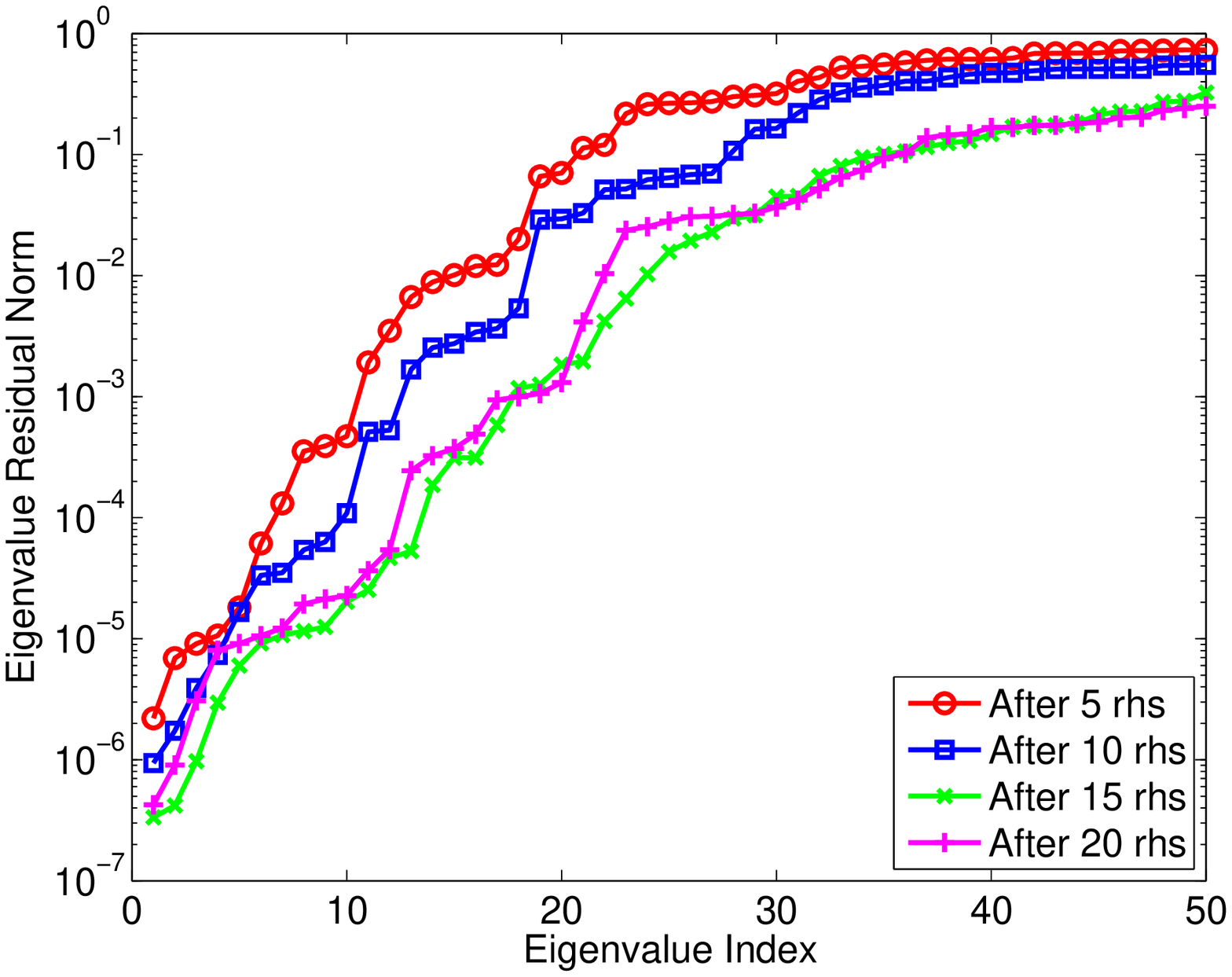}
\includegraphics[width=0.45\textwidth ]{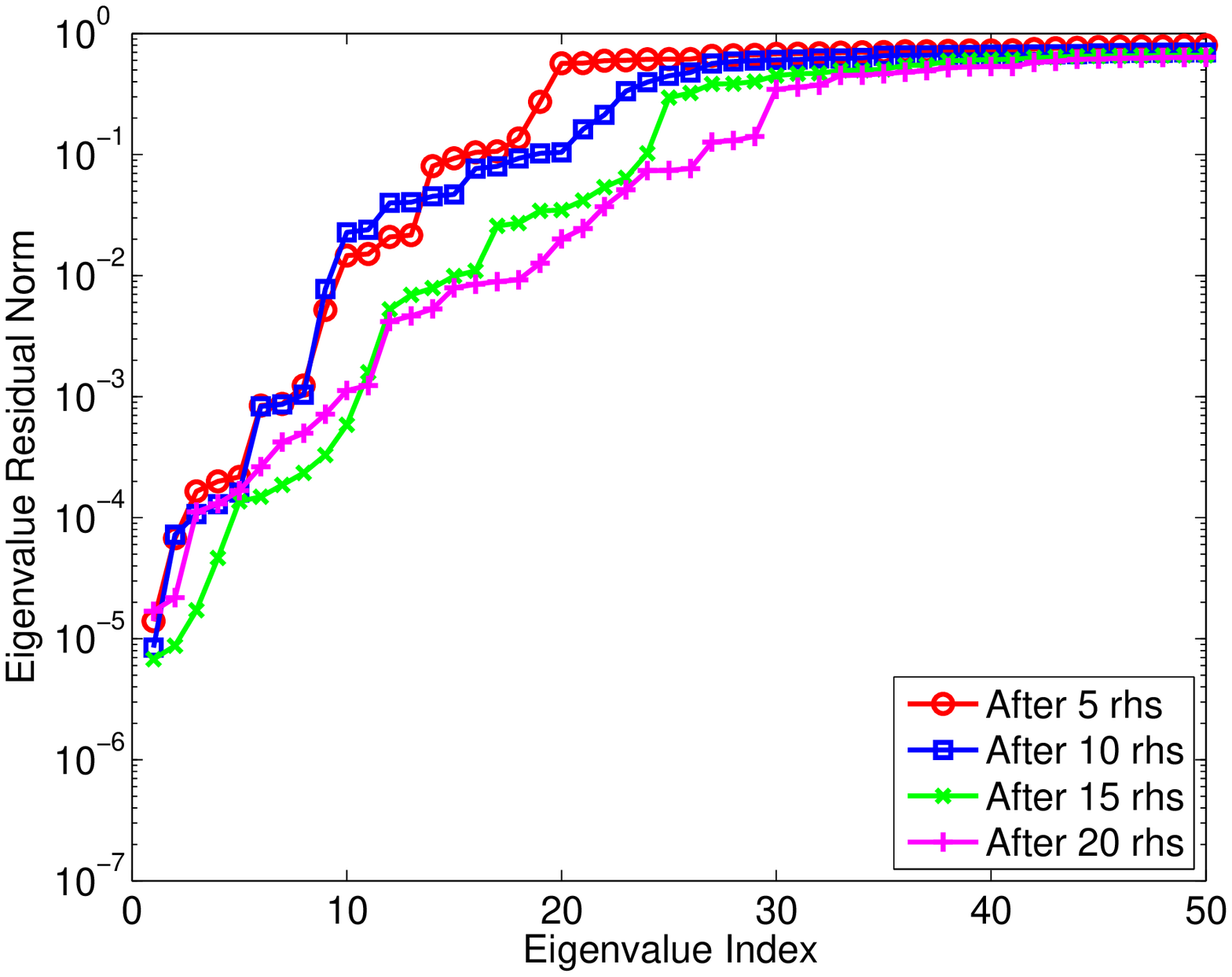}
\caption{Improvement of the accuracy of the best 50 eigenvalues computed with Incremental eigBiCG(15,40) as more 
systems are solved for the matrices  $QCD-49K$(left) and $QCD-249K$(right).}
\label{Fig:qcd-eval-incremental-improvement}
\end{center}
\end{figure}

We next show results for the QCD matrices.
For these tests we used $m=40,$  $k=15$, $tol=10^{-10}$,
and $btol=10^{-4}$. \initbicgstab was only 
restarted once when the linear system converged to a tolerance of $10^{-8}$.
In Figure \ref{Fig:qcd-defbstab-vs-nrhs}, we compare \bicgstab to \initbicgstab 
where the number of deflated eigenvectors is obtained from different numbers 
of right hand sides. Overall, just a few eigenvectors yield a speedup of two 
or more.
To illustrate the improvement of the eigenvectors as we solve more systems, 
we show in Figure \ref{Fig:qcd-eval-incremental-improvement} the residual 
norm for the best 50 eigenvalues computed and how this improves over time.

We conclude this subsection by observing that in all our 
previous experiments, a single restart of the deflated \initbicgstab gave
the best convergence.
Therefore, as long as the vectors can be stored, the computational cost of 
applying the deflation 
projector is negligible (in QCD problems one matrix-vector operation costs about the same as an application of 
a projector with 300 vectors).

\subsection{Comparing with GMRes--DR/GMRes--Proj}
The {\it GMRes-DR(m,k)} algorithm \cite{Morgan:2002} solves a nonsymmetric
linear system using restarted {\it GMRes} and simultaneously computes $k$ approximate eigenvectors.
Like \eigbicg, it uses a subspace of maximum size $m$ which is restarted to update
$k$ approximations to the desired eigenvectors.  
Unlike \eigbicg, however, it explicitly orthogonalizes future iterates to these $k$ eigenvector 
approximations, thus improving also the convergence of the restarted {\it GMRes(m)}.
In theory, the advantages of \eigbicg are that 
(a) the biorthogonality of the whole space is implicit,
(b) it uses not only thick but also locally optimal restarting to update the $k$ eigenvectors,
(c) the underlying Krylov method is unrestarted, and
(d) produces both left and right eigenvectors.
The advantage of {\it GMRes-DR(m,k)} is that it is equivalent to the IRA eigensolver \cite{ARPACK}.
In practice, the most important difference is the performance of the underlying methods
({\it GMRes(m)}, \bicg~) on a particular problem.

For systems with multiple right-hand sides, the computed eigenvectors from the first system 
are used to deflate {\it Restarted GMRes} for the following systems. 
Because it is expensive to deflate these $k$ vectors at every step of {\it GMRes-DR(m,k)}, 
they are used in the {\it GMRes-Proj} method \cite{Morgan:2004}. 
In {\it GMRes-Proj}, cycles of {\it GMRes($m'$)} are alternated with a minimum residual projection 
over these $k$ eigenvectors. To maintain the same memory cost, usually $m'=m-k$. 
Therefore, {\it GMRes-Proj} applies deflation only periodically, like our 
restarted \initbicgstab.
The difference is that {\it GMRes-Proj} applies the projection every $m'$ steps and thus the total number
of projections depends on the convergence rate of the problem, while 
\initbicgstab is restarted a constant number of times, $tol/rtol$.
Moreover, all eigenspace information comes from one run of {\it GMRes-DR(m,k)}, while
 \increigbicg builds the eigenspace by accumulating vectors from $n_1$ right-hand sides.

A thorough comparison between \increigbicg and {\it GMRes--DR}/{\it GMRes-Proj} requires
experimentation on a large parametric space, with different objectives (time, memory, iterations), 
and application problems. This is beyond the scope of this paper. 
Instead, we provide a sample experiment that shows that our method is competitive to a 
state-of-the-art method for solving systems with multiple right-hand sides.
We use the two QCD matrices from our previous experiments and report also 
timings because the methods have different costs per iteration.  

We solve linear systems for 100 random right-hand sides to $||r|| < 10^{-10}||b||$.
After solving the first system with {\it GMRes--DR(80,60)}, we obtain 60 (approximate) eigenvectors which 
we deflate at every cycle of {\it GMRes(20)-Proj(60)} for the next 99 systems.
For \increigbicg, we solve the first 5 systems using {\it eigBiCG(12,40)} accumulating 60 left and 
right eigenvectors. These are then used to deflate \initbicgstab without restarting for the rest 95 systems.
To match the memory used by \increigbicg, we also compare against 
{\it GMRes--DR(140,120)} followed by {\it GMRes(20)--Proj(120)}. The large subspace makes the latter 
method more expensive per step but it should have better deflation properties.

\begin{figure}[htbp]
\begin{center}
\includegraphics[width=0.4475\textwidth]{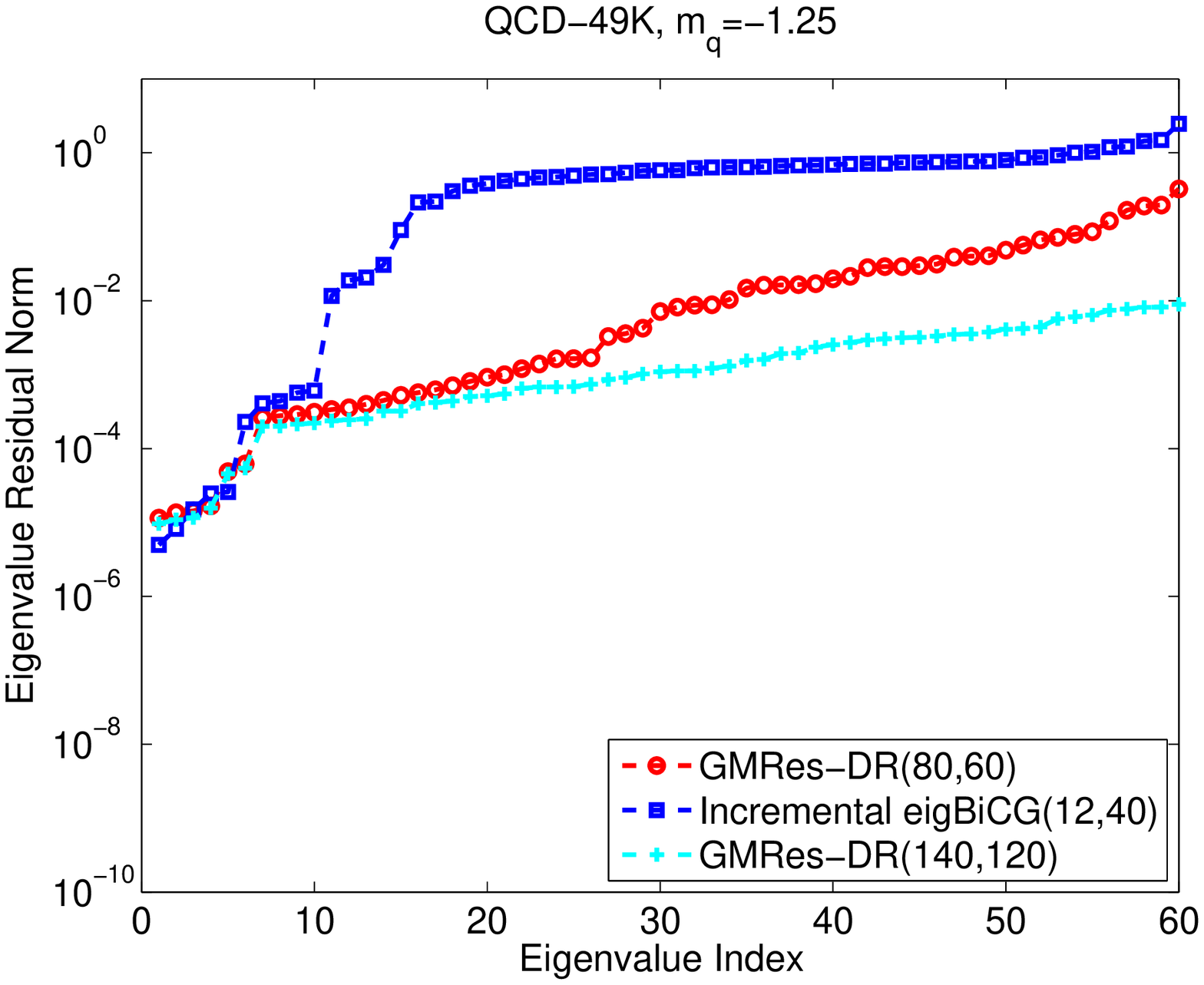}
\includegraphics[width=0.4525\textwidth]{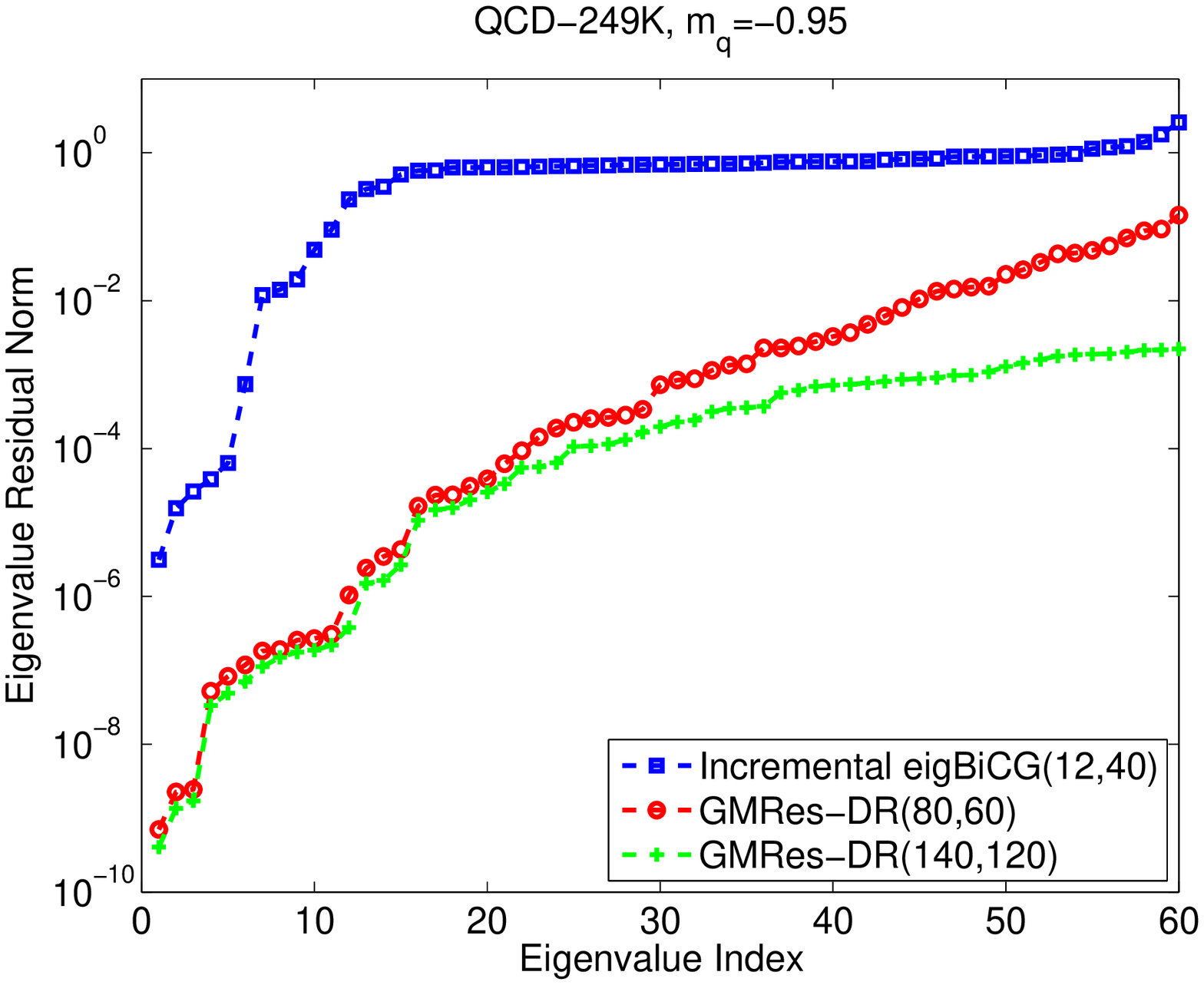}
\caption{Residual norms of the lowest 60 eigenvalues of the QCD matrices computed using GMRes--DR(60,80),
GMRes--DR(120,140), and Incremental eigBiCG(12,40) for 5 right-hand sides.}
\label{Fig:qcd49k-249k-comp-eres-gdr-eigbicg}
\end{center}
\end{figure}

In Figure \ref{Fig:qcd49k-249k-comp-eres-gdr-eigbicg}, we compare the residual norms of the 
best 60 eigenvectors computed by each of the three methods. 
We mention that the eigenvalues of the QCD matrices are symmetrically located around 0 which does not favor \bilanczos.
As an exact eigensolver with a large subspace (80 or 140 vectors) 
{\it GMRes-DR} produces better residual norms than \increigbicg. 

\begin{figure}[htbp]
\begin{center}
\includegraphics[width=0.4535\textwidth]{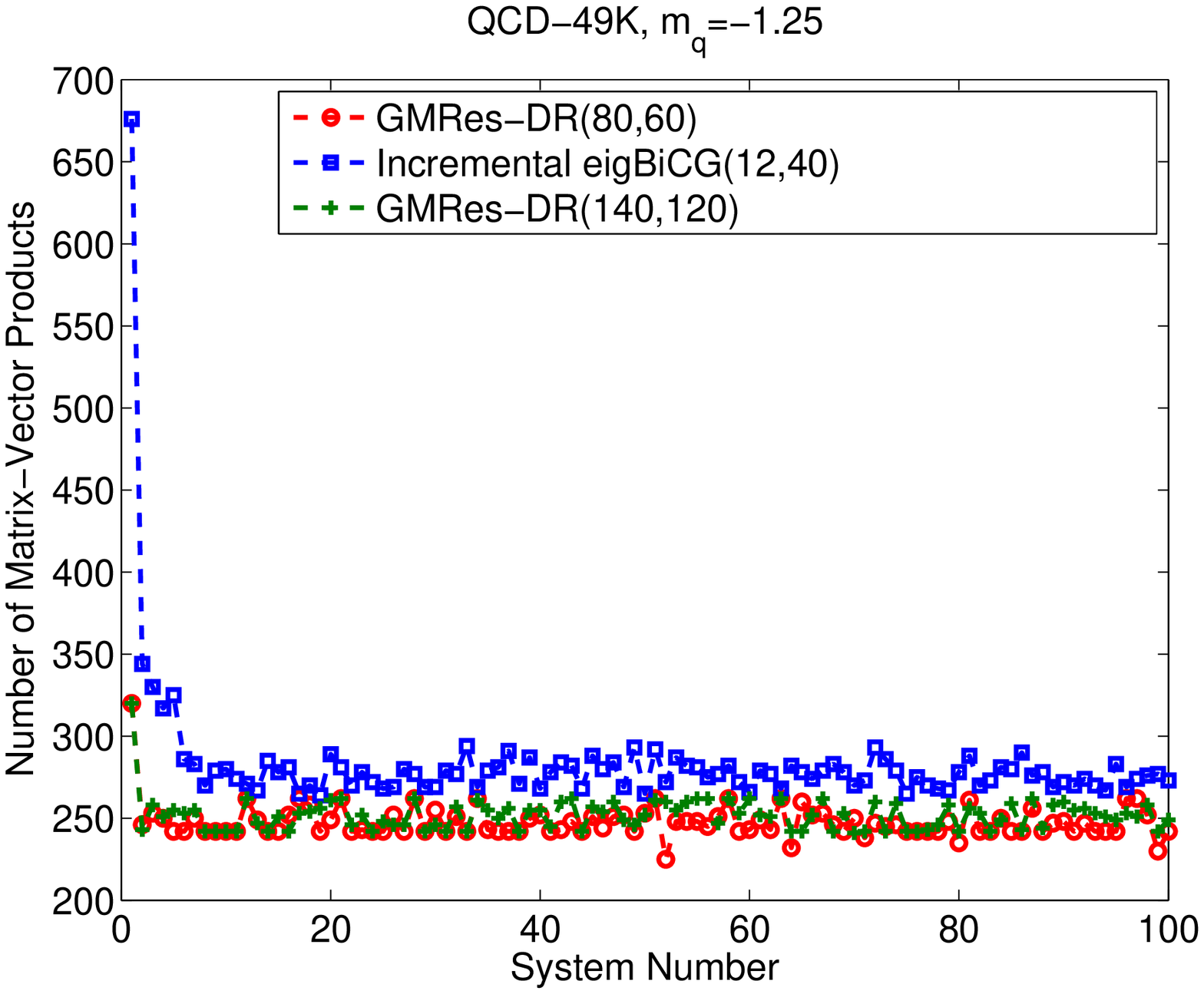}
\includegraphics[width=0.4465\textwidth]{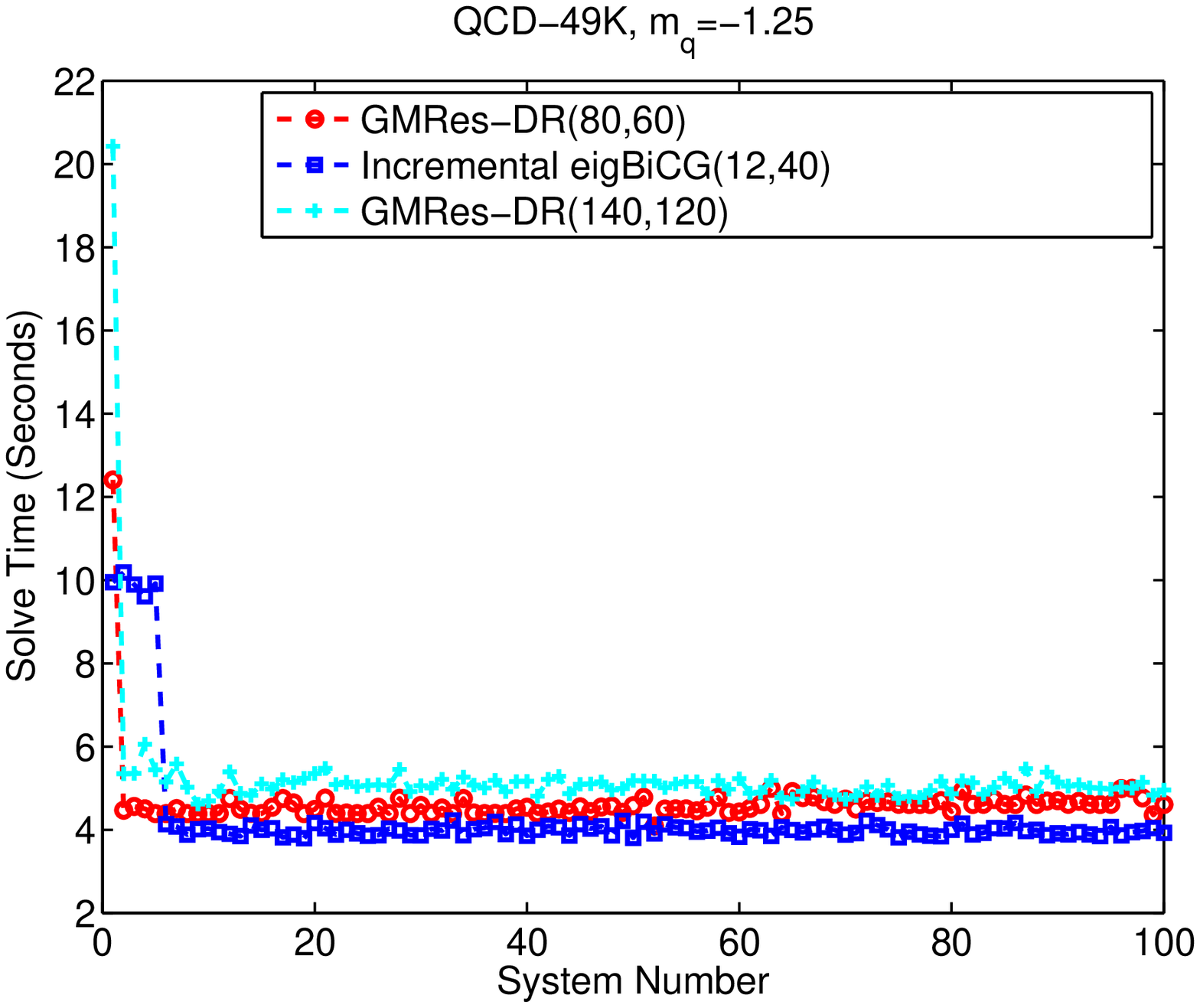}
\caption{Solving 100 right-hand sides using GMRES--DR(m,k) and Incremental eigBiCG(nev,m)
for the QCD--49K matrix. The first system is solved with GMRES--DR(m,k) and the 
subsequent 99 systems are solved using GMRES(m-k)--Proj(k) in which k eigenvectors are deflated. For 
Incremental eigBiCG(nev,m), the first 5 systems are solved with eigBiCG(nev,m) and the subsequent 95 systems 
with init-BiCGStab with 5*nev eigenvectors deflated. On the left, we show the number of matrix-vector
products in both cases. On the right we show the solution time. For this problem, Incremental eigBiCG is faster
than GMRES--DR.}
\label{Fig:qcd49k-comp-linsys-gdr-eigbicg}
\end{center}
\end{figure}

Figure \ref{Fig:qcd49k-comp-linsys-gdr-eigbicg} shows the cost for solving each of the 100 systems
for {\it QCD--49K}. 
For the first system the number of iterations is similar for all methods, but \bicg requires two 
matrix-vector products per iteration. For subsequent deflated systems, \bicgstab required only a few 
more products than the {\it GMRES--Proj} variants. The right part of the figure shows that 
the inexpensive deflation and iteration step of \initbicgstab make it faster than {\it GMRES--Proj}, 
especially when a large number of right-hand sides need to be solved.
The only exception is the short incremental phase where \bicg is used which converges slower than \bicgstab.

\begin{figure}[htbp]
\begin{center}
\includegraphics[width=0.454\textwidth]{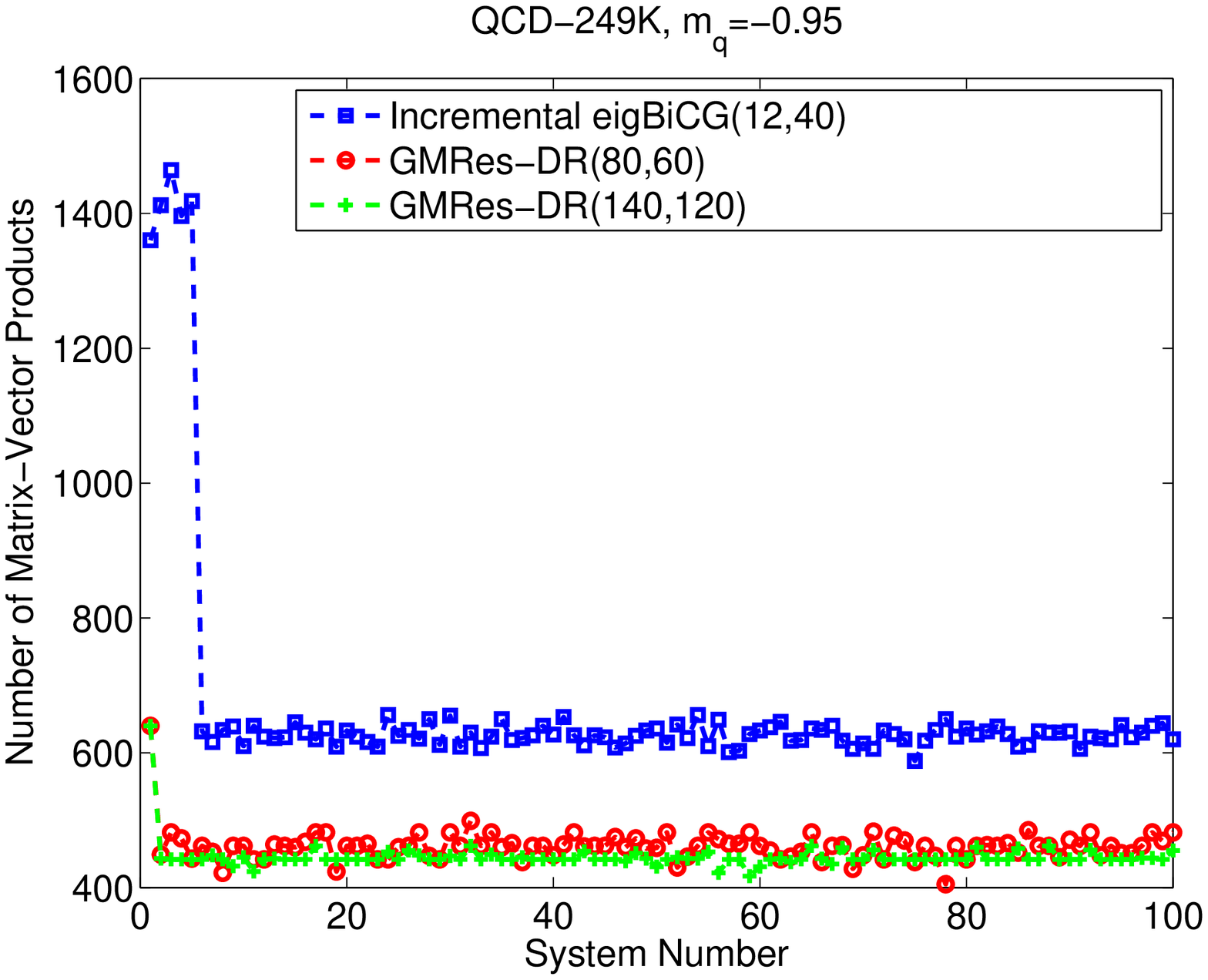}
\includegraphics[width=0.446\textwidth]{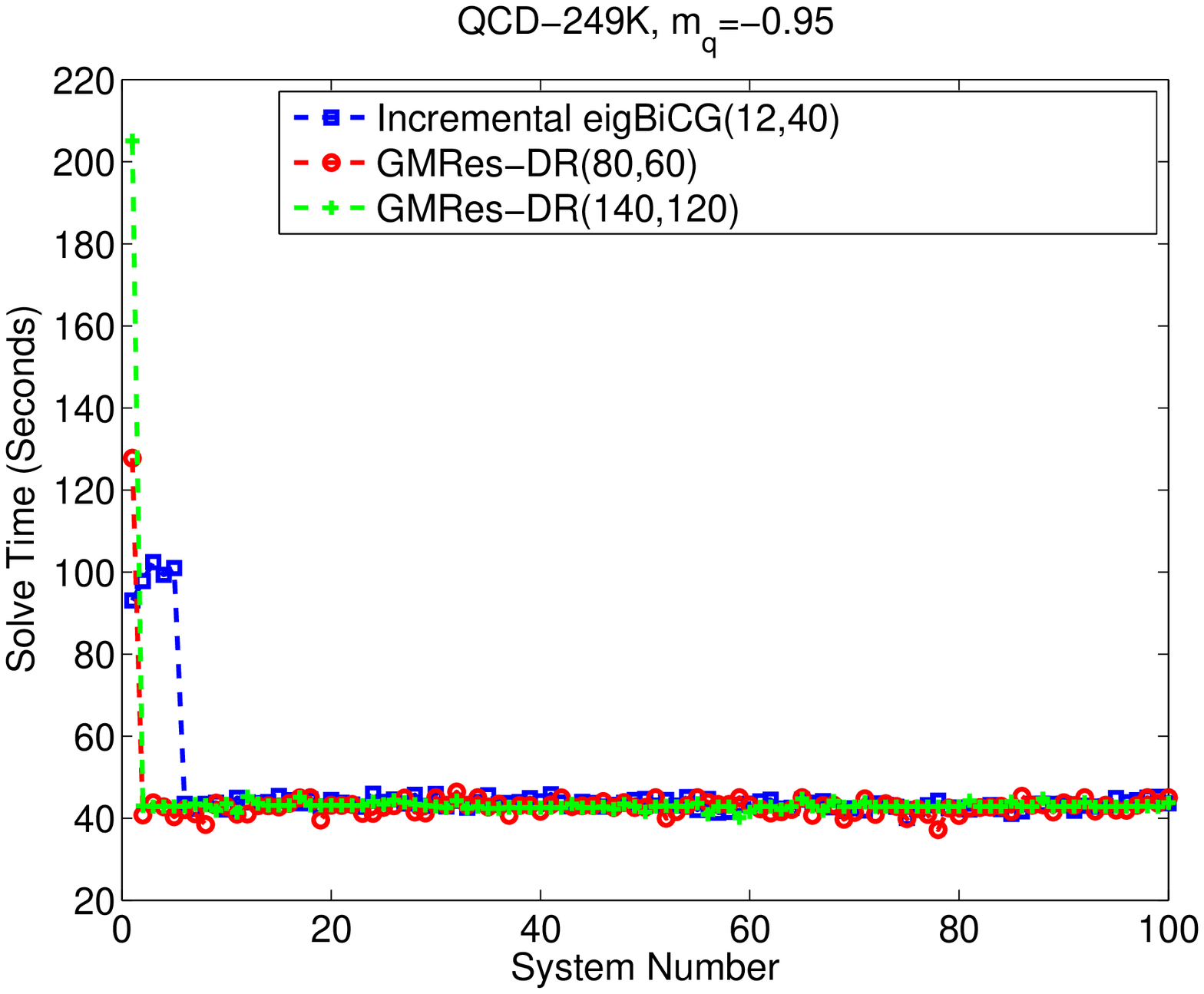}

\caption{
Solving 100 right-hand sides using GMRES--DR(m,k) and Incremental eigBiCG(nev,m)
for the QCD--249K matrix. The procedure and parameters are the same as in Figure 
 \ref{Fig:qcd49k-comp-linsys-gdr-eigbicg}.
On the right we show the solution time. For this problem, Incremental eigBiCG is equally fast to GMRES--DR.}
\label{Fig:qcd249k-comp-linsys-gdr-eigbicg}
\end{center}
\end{figure}

Figure \ref{Fig:qcd249k-comp-linsys-gdr-eigbicg} shows similar results for the 
matrix {\it QCD--249K}. \initbicgstab took about 50\% more matrix-vector products than {\it GMRes-DR}
(although the number of iterations was smaller) but all methods achieved solutions in similar times.

We note that the parameter choices for \increigbicg were not the best ones identified in previous sections 
because we wanted all methods to use either the same number of deflation vectors or the same memory.
With the best parameters, the number of matrix-vector products of \increigbicg is less than that of
{\it GMRES--Proj} for {\it QCD--49K} and about the same for {\it QCD--249K}
(see Figure~\ref{Fig:qcd-defbstab-vs-nrhs}) and thus we expect our method to be quite faster. 

\section{Conclusions}
\label{sec:conclusions}
We have extended the \eigcg algorithm for solving linear systems with multiple right-hand sides
to the nonsymmetric case.
The resulting algorithm, \eigbicg, approximates a few smallest magnitude eigenvalues and their
corresponding left and right eigenvectors while a linear system is solved with \bicg.
The algorithm uses only a small size window of the \bicg residuals without affecting the 
convergence of the linear system and without restarting \bicg. The \eigbicg algorithm was tested on matrices from different applications.
For nonsymmetric, non-defective matrices with a positive definite symmetric part, \eigbicg
was able to compute eigenvalues almost as accurately as those computed with unrestarted and even
explicitly biorthogonalized \bilanczos algorithms. 

For systems with multiple right-hand sides, we have given an algorithm that incrementally improves the number and accuracy of the
eigenvalues computed with \eigbicg while solving the first few systems. The computed eigenvectors are then used to deflate
\bicgstab not at every step, but only initially at the right-hand side. Repeating this deflation once or twice
by restarting \bicgstab was always sufficient.
In our experiments our deflated method achieved speedups of a factor of two or 
more.
We also showed that the method is competitive to a state-of-the-art method for multiple right-hand sides,
the {\it GMRes-DR/GMRes-Proj}.

Further improvements of the algorithms that are also relevant for the SPD case could be investigated in the future. 
Examples include, how to implement selective biorthogonalization to reduce the effect of biorthogonality loss
in \eigbicg, how to reduce the number of accumulated vectors in \increigbicg by restarting the bases, 
or what the effect of deflation is on the accuracy of the solution 
of the linear system.

\acks
This work was supported by the National Science Foundation grant CCF-0728915, the DOE Jefferson Lab, the Jeffress 
Memorial Trust grant J-813, and received partial support through the Scientific Discovery through Advanced 
Computing (SciDAC) program funded by U.S. Department of Energy, Office of Science, Advanced Scientific Computing 
Research and Nuclear Physics under award number DE-FC02-12ER41890. A. Abdel-Rehim was partially supported
by the PRACE-2IP project (Partnership for Advanced Computing in Europe, Second Implementation Phase) under WP8 
grant number EC-RI-283493.

\bibliography{refs}
\bibliographystyle{wileyj}
\end{document}